\documentclass[apj]{emulateapj}

\usepackage{graphicx}
\usepackage{calrsfs}
\usepackage{amsbsy}
\usepackage{amssymb}
\usepackage{amsmath}

\usepackage[hypertex]{hyperref}

\begin{document}

\def\be{\begin{equation}}
\def\ee{\end{equation}}
\def\ba{\begin{eqnarray}}
\def\ea{\end{eqnarray}}
\def\d{\delta}
\def\e{\epsilon}
\def\f{\varphi}
\def\t{\tilde}
\def\p{\partial}
\def\ms{\mathstrut}
\def\s{\strut}
\def\ds{\displaystyle}
\def\ts{\textstyle}
\def\b{\boldsymbol}
\def\r{\mathrm}
\def\k{\varkappa}
\def\vt{\vartheta}
\def\mnras{MNRAS}
\def\jgr{JGR}
\def\prd{Phys. Rev. D}
\def\nat{NAT}
\defcitealias{Medved00}{M00}
\defcitealias{Toptygin_Fleish}{TF87}


\title{On the Jitter Radiation}

\author{S.R.~Kelner}
\affil{National Research Nuclear University (MEPHI), Kashirskoe shosse
31, 115409 Moscow, Russia; Max-Planck-Institut f\"ur Kernphysik,
Saupfercheckweg 1, D-6917 Heidelberg, Germany}
\email{skelner@rambler.ru}

\author{F.A.~Aharonian}
\affil{Dublin Institute for Advanced Studies, 31 Fitzwilliam Place, Dublin 2, Ireland;
Max-Planck-Institut f\"ur Kernphysik,
Saupfercheckweg 1, D-6917 Heidelberg, Germany}
\email{Felix.Aharonian@mpi-hd.mpg.de}

\author{D.~Khangulyan}
\affil{Institute of Space and Astronautical Science/JAXA, \\
3-1-1 Yoshinodai, Chuo-ku, Sagamihara, Kanagawa 252-5210, Japan}
\email{khangul@astro.isas.jaxa.jp}

\begin{abstract}
In a small scale turbulent medium, when the nonrelativistic Larmor radius $R_{\rm L}=mc^2/eB$ exceeds the
correlation length $\lambda$ of the magnetic field, the magnetic
bremsstrahlung of charged relativistic particles unavoidably proceeds in the
so-called jitter radiation regime. The cooling timescale of parent particles
is identical to the synchrotron cooling time, thus this radiation regime can
be produced with very high efficiency in different astrophysical sources
characterized by high turbulence. The jitter radiation has distinct spectral
features shifted, compared to synchrotron radiation, towards high energies.
This makes the jitter mechanism an attractive broad-band gamma-ray production
channel which in highly magnetized and turbulent environments can compete or
even dominate over other high energy radiation mechanisms. In this paper we
present a novel study on spectral properties of the jitter radiation performed
within the framework of perturbation theory. The derived general expression
for the spectral power of radiation is presented in a compact and convenient for
numerical calculations form.

\end{abstract}


\maketitle

\section{Introduction}

Charged particles moving in electric and magnetic fields experience effective
energy losses via radiation. Because of high conductivity, the electric fields
in astrophysical plasmas are typically screened, thus the
radiation is dominated by interactions with the magnetic field due to the
so-called {\it magnetic bresstrahlung}. The latter is one of the major
nonthermal radiation processes in astrophysics and operates with high
efficiency in a large variety of astrophysical environments. In the case of a
regular magnetic field or a chaotic field characterized by large scale
fluctuations, we deal with the so-called synchrotron radiation. This process and
its implications in astrophysics have been studied in great detail \citep[see
e.g.][]{Ginz_Syr, Rybicki}. In highly turbulent environments, namely, when the
nonrelativistic Larmor radius $R_{\rm L}=mc^2/eB$ does not exceed the characteristic scale of
turbulence $\lambda$, the radiation proceeds in significantly different regime,
which in the astrophysical literature is referred as {\it diffusive synchrotron
radiation} \citep[][hereafter \citetalias{Toptygin_Fleish}]{Toptygin_Fleish} or as {\it jitter
radiation} \citep[][hereafter \citetalias{Medved00}]{Medved00}. Hereafter we will use the term
``jitter''.

The spectral features of jitter radiation substantially differ from
the synchrotron radiation. While the power of the synchrotron
radiation of a monoenergetic particle $P_\omega$ is described with a
good accuracy as $\omega P_\omega \propto \omega^{4/3}
\exp[-(\omega/\omega_c)]$, where $\omega_c=3 \gamma^2eB/(2mc)$ is the
characteristic synchrotron frequency, in the case of jitter radiation
the peak is shifted towards higher frequencies by a factor of
$a=R_{\rm L}/\lambda$. Unless the distribution of emitting particles
is strictly monodirectional, the power spectrum of jitter radiation below the
maximum is flat, i.e. the Spectral Energy Distribution
(SED)\footnote{The so-called Spectral Energy Distribution or SED is
  determined as $\omega^2 dN/d\omega$ or $\epsilon^2 dN/d\epsilon$,
  where $dN/d\omega$ ($dN/d\epsilon$) is the differential distribution
  over frequencies (energies). Obviously here the SED is $\omega
  P_\omega$; note that in astronomical literature for SED is often
  used the denotation $\nu F_\nu$.}  $\omega P_\omega \propto
\omega_{}$, while beyond the cutoff energy it has a power-law
behavior, $\omega P_\omega \propto \omega^{1-\alpha}$, where $\alpha$
is the power-law index of the turbulence spectrum
\citepalias{Toptygin_Fleish}. Thus, instead of the typical exponential
cutoff in synchrotron spectrum, the jitter mechanism yields a
power-law spectrum which can be extended up to the frequency of $a^3
\omega_c$. This makes the jitter radiation of electrons an excellent
{\it high energy} gamma-ray production process in contrast to the
synchrotron radiation which even in the case of extreme accelerators
operating at the maximum possible rate allowed by classical
electrodynamics \citep{Aharonian2002} is limited by the maximum
possible energy $\epsilon_0=\hbar \omega_0=9/4 \alpha^{-1} mc^2 \sim
150 \ \rm MeV$.

However, so far this remarkable feature of jitter radiation
practically has not been explored for interpretation of {\it high
  energy} gamma-ray phenomena \citep[see however, the recent papers by
][]{teraki_takahara_13}. Instead, more emphasis has been
placed on the energy interval below the cutoff. In particular,
it is augured in \citetalias{Medved00} that the jitter radiation below the
cutoff can result in harder spectra than the synchrotron radiation,
namely $\omega P_\omega \propto \omega^2$. However, the claimed
energy dependence is closely related to the assumed geometry of the
magnetic field. Namely, it can be achieved if the magnetic field has
only one non-zero component which  can be realized only for a rather unrealistic
configuration of the turbulent field (see Sect.~\ref{compare} for details).

In the seminal paper on jitter radiation by
\citetalias{Toptygin_Fleish} it has been realized that the spectral
maximum of the jitter emission is located at higher frequencies than
in the synchrotron regime, and that the high energy part of the jitter
spectrum could be described by a power-law. Thus, even for the case of
monoenergetic particle distribution one may expect a broken power-law
spectrum. This should lead to the modification of the standard
relations between spectral slopes, flux levels and breaks found in
synchrotron spectra.  Possible applications of the jitter mechanism
also has been discussed, basically in the low energy range of cosmic
electromagnetic radiation.  In particular, it has been proposed that
the jitter radiation can be responsible for the radio to optical
(X-ray) spectra of some  active galaxies and pulsar wind
nebulae (\citetalias{Toptygin_Fleish};\citealt{Fleish_Biet,mao2007}).

The underestimation of the potential of jitter radiation for production of
high and very high energy
gamma-rays  could be related to the effect of weakening of the
diffusive shock acceleration process in the case of short-length scale
turbulence. A self-consistent consideration of the processes of particle
acceleration and emission in the framework of the diffusive shock acceleration
paradigm predicts a shift of the
jitter radiation peak towards low frequencies as compared to the pure
synchrotron radiation \citep{derishev07,kirk_reville10}. However, if the
inhomogeneities responsible for particle acceleration and emission are
different, e.g. when these processes occur in spatially separated regions,
the spectral maximum would be shifted towards higher energies making the
jitter radiation a very effective high energy gamma-ray production mechanism.
Therefore the spectral features of this radiation in the entire energy range
deserve detailed qualitative studies.

To explore the process in a general form, we propose a new approach based on
the perturbation theory. In terms of additional assumptions, the proposed
method is less demanding compared to previous studies, and allows a precise
control of the applicability conditions for the derived solutions, e.g. the
range of the high energy power-law extension beyond the spectral maximum.

In this regard we should note that in previous studies some principal results
have been obtained under specific, although not always obvious assumptions.
For example, \citetalias{Medved00} has derived the spectrum of radiation for the
case of a very specific geometry of the magnetic field fluctuations. In some
others studies \citep[see e.g.,][]{Fleish_Biet} the jitter radiation spectrum
in fact has not been strictly derived, but rather {\it predefined} through
its asymptotic behavior. Finally, some studies address the case of anisotropic
turbulence \citep{Medved12}, however the structure of the used correlation
tensor
is not consistent with the fundamental requirement of $\nabla\cdot \b B=0$. We
discuss these concerns in detail in Section~\ref{compare}.

Finally,  we should mention that a significant progress recently has been achieved through numerical computations 
based on particle-in-cell technique \cite[see, e.g.,][]{reville2010,teraki_takahara_11}.  This method has a great potential 
to deal with quite complex distributions of emitting particles. On the other had,  the analytical approach allows better 
understanding and interpretation of physics behind   the obtained results. In this regard, two methods  
are complementary and equally  important.

The paper is organized as following: in Section \ref{sec:perturb} the basic
results on the energy
spectra, as well as the applicability limits for the derived spectra are
presented. in Section \ref{sec:chaotic} we consider the case of chaotic
magnetic field. In Section \ref{sec:comp} we compare the radiation properties
in chaotic magnetic field with the conventional synchrotron radiation; the
latter is is briefly discussed in Section \ref{sec:synch}. The case of
anisotropic turbulence (under assumption of isotropic distribution of emitting
particles) is considered in Section \ref{isotrop}. Finally, we compare our
results with previous studies in Section \ref{sec:prev}, and summarize the
main results in Section \ref{sec:conc}.

\section{Perturbation Theory}\label{sec:perturb}
The intensity and the energy distribution of radiation produced by a particle
of a given charge $e$
depends only on its trajectory. Let $\b r(t)$ and $\b v(t)=\dot{\b r}(t)$
be the radius-vector and the velocity
of the particle at the instant $t$. Then, the energy spectrum of radiation
is described by
equation (14.65) of \citet{Jackson} which for our purposes is convenient to
present in the
form\footnote{$\rm d\mathcal{E}_{\b n\omega}$ is the energy radiated by a particle
into the solid angle $\rm d\Omega$ within the frequency interval $\rm d
\omega$.}
\be\label{jitt_1}
\begin{split}
\frac{d\mathcal{E}_{\b
n\omega}}{d\omega\,d\Omega}=&\frac{e^2}{4\pi^2c^3}\left|\int\limits_{\mathbb{R}}
^{} \b U(t)\,dt\right|^2\\
=&\frac{e^2}{4\pi^2c^3}\int\limits_{\mathbb{R}^2}^{} \b U(t_1)\,\b
U^*(t_2)\,dt_1\,dt_2\,.
\end{split}
\ee
Here the integrand
\be\label{jitt0}
\b U(t)=\frac{\b n\times[(\b n-\b\beta(t))
\times\b a(t)]}{(1-\b n\b\beta(t))^2}\,e^{i\Phi(t)}
\ee
depends on the particle velocity $\b v(t)=c\b\beta(t)$ and the acceleration
$\b a(t)=c\dot{\b\beta\,}\!(t)$, as well as the function $\Phi(t)=\omega(t-\b
n\b r(t)/c)$, where $\b n$ is the unit vector towards the momentum of the
radiated photon. The function $\b U^*(t)$ is the is complex conjugation of
$\b U(t)$. Note that equation \eqref{jitt_1} is precise; it is derived within
the framework of classical electrodynamics through integration of the Maxwell
equations in vacuum.

It is convenient to introduce new variables of integration:
$t=(t_1+t_2)/2$ and $\tau=t_1-t_2$ (note that $dt_1\,dt_2=dt\,d\tau$). Then
equation~(\ref{jitt_1}) results in
\be\label{jitt1}
\frac{d\mathcal{E}_{\b n\omega}}{d\omega\,d\Omega}=
\frac{e^2}{4\pi^2c^3}\int \b U(t+\tau/2)\,\b U^*(t-\tau/2)\,dt\,d\tau\,.
\ee
As it is shown bellow, for a fixed value of $t$ the integrand rapidly decreases
with the increase of $|\tau|$. Also,
the integrand is characterized by a weak dependence on $t$. The integration of
the integrand over $\rm d \tau$
gives the {\it spectral power} of emission at the moment $t$:
\be\label{jitt2}
P_{\b n\omega}(t)=\frac{e^2}{4\pi^2c^3} \int\limits_{-\infty}^{\infty}
\b U(t+\tau/2)\,\b U^*(t-\tau/2) \,d\tau\,.
\ee
Note that if one considers equation~(\ref{jitt1})
as a classical limit of the corresponding quantum relation, then the integrand
in equation~(\ref{jitt1})
can be interpreted as the emission probability multiplied to the photon energy
 $\hbar\omega$
\citep[see][\S~90]{Berest}.

The radiation detected by an observer is produced by ensemble of
particles occupying a certain region in space. We will consider the
case of a chaotic magnetic field, assuming that statistically averaged (over time or space)
magnetic field $\langle\b B\rangle =0$. 

Let $\lambda$ be the
correlation length of the magnetic field.
If the distance between two chosen points at $\b r_{1}$ and $\b r_{2}$ exceeds
$\lambda$, then the corresponding magnetic fields $\b B_1$
and $\b B_2$ can be treated as statistically independent, thus the
time-averaged
product of these fields $\langle B_{1\rho} B_{2\sigma}\rangle=\langle
B_{1\rho}\rangle
\langle B_{2\sigma}\rangle=0$.

To obtain the radiation spectrum, the integrand in equation~(\ref{jitt2})
should be averaged over
all possible configurations of the magnetic field. It is convenient to perform
this procedure in the framework of perturbation theory. The acceleration of
particle
is proportional to the strength of the magnetic field $\b B$, $\b a=e(\b\beta
\times \b B)/(m\gamma)$.
In the first approximation, all other relevant parameters can be treated as in
the absence of the magnetic field, i.e.
$\b\beta(t\pm\tau/2)=\b\beta(t)$, $\b r(t\pm\tau/2)=\b r(t)\pm\b\beta(t)\tau/2$.
The applicability of the approach is discussed bellow. This approximation
results in
\[
P_{\b n\omega}(t)= \frac{e^2}{4\pi^2c^3 (1-\b n\b\beta)^2}
\]
\be\label{jtt4}
\times\int\limits_{-\infty}^{\infty} \left[ \b a_+\b a_- -
\frac{(\b n\b a_+)(\b n\b a_-)}{\gamma^2(1-\b n\b\beta)^2} \right]
 e^{i\omega(1-\b n\b\beta)\,\tau}\,d\tau\,,
\ee
where $\b a_{\pm}=\b a(t\pm\tau/2)$ and $\b \beta=\b \beta(t)$);
$\gamma=1/\sqrt{1-\beta^2}$ is the particle Lorentz factor.
The derivation of equation~(\ref{jtt4}) was performed using the formula for
the
standard double vector product:
$\b a\times(\b b\times\b c)={\b b}(\b a\b c)-{\b c}(\b a\b b)$ and taking into
account that in the magnetic field
the acceleration and velocity vectors are orthogonal.
Equation~(\ref{jtt4}) represents the first non-vanishing term in the expansion
of the emission spectrum
 in powers of the magnetic field.

Our ultimate aim is to derive the emission spectrum integrated over the
emission angles of photons and averaged over the magnetic field fluctuations.
It is convenient to select the $z$ axis to be parallel to the particle
velocity $\b\beta$, and start
with averaging over the azimuthal angle $\phi$
in respect to the direction of the particle velocity $\b\beta$. Then, the
scalar product of the
vectors $\b n$ and $\b \beta$ does not depend on the azimuthal angle $\phi$,
$(\b n\b\beta)=\beta\cos\theta$.
Given that $\b a_{\pm}\perp \b\beta$, one obtains
\be\label{jtt6}
\langle(\b n\b a_+)(\b n\b a_-) \rangle=\frac12\,(\b a_+\b a_-) \sin^2\theta\,.
\ee
and, after averaging of equation~(\ref{jtt4}) over $\phi$,
we have
\[
P_{\b n\omega}(t)= \frac{e^2}{4\pi^2c^3
(1-\b n\b\beta)^2} \left( 1- \frac{\sin^2\theta}{2\gamma^2(1-\b n\b\beta)^2}
\right)
\]
\be\label{jtt7}
\times \int\limits_{-\infty}^\infty (\b a_+\b a_-)\,
e^{i\omega(1-\b n\b \beta)\tau}d\tau\,.
\ee

\begin{figure}
\centering{
\includegraphics[width=0.36\textwidth,angle=0]{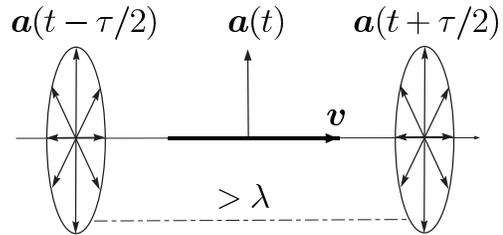}
}
\caption{Schematic description of the basic geometry adopted for computations.
Thin solid line
 represents a segment of the trajectory of the charge particle, which in the
framework of the
 perturbation approach can be taken as a straight line. The
 particle acceleration is orthogonal to the velocity (which in its turn is
parallel to the trajectory). Also it is assumed that the acceleration is
statistically independent for distances along the trajectory
 which exceed the magnetic field correlation length $\lambda$. This corresponds
to the condition
 $|\tau|>\lambda/c$.
}
\label{fig:jitt_a3}
\end{figure}

In equation~(\ref{jtt7}) the charge velocity, $\b \beta=\b\beta(t)$, is treated
as a constant (independent of $\tau$).
The averaging over the magnetic field configurations results in appearance of
a correlation function,
$\langle \b a_+a_-\rangle$, under the integral. Note that for $\b \beta= \rm
const$, the acceleration and magnetic field have the same statistical
properties. In particular, $\langle \b a(t)\rangle=0$ and $\langle \b
a(t+\tau/2)\,\b a(t-\tau/2)\rangle= \langle \b a(t+\tau/2) \rangle \langle \b
a(t-\tau/2) \rangle=0$ if the distance between the corresponding points exceeds
$\lambda$ (i.e., if $c\beta\tau>\lambda$). This feature of $\langle \b
a_+a_-\rangle$ is illustrated in figure~\ref{fig:jitt_a3}.

Since the radiation of ultrarelativistic particles is strongly beamed
towards the direction of motion
($\theta \sim 1/\gamma$), we will consider only the region of small angles,
$\theta \ll1$. This allows significant simplifications of
calculations which result in
\be\label{jtt8}
P_{\b n\omega}(t)= \frac{e^2}{\pi^2c^3}\,
\frac{\gamma^4(1+\gamma^4\theta^4)}{(1+\gamma^2\theta^2)^4}
\int\limits_{-\infty}^\infty \langle\b a_+\b a_-\rangle\,e^{i\t\omega
\tau}\,d\tau\,,
\ee
where
\be\label{jtt9}
\t\omega=\frac{\omega}{2\gamma^2}\,(1+\gamma^2\theta^2)\,.
\ee

Since the above results are derived within the framework of the perturbation
approach,
it is important to study the range of applicability of equation~(\ref{jtt8}).

The integrand in equation~(\ref{jtt8}) rapidly decreases in the range of
$|\tau|\gtrsim \lambda/c$.
Therefore, the obtained expression describes correctly the emission power
if the terms neglected at derivation of equation~(\ref{jtt8}) are small for
$|\tau|\lesssim\lambda/c$.
In the precise equation~(\ref{jitt2}), the denominator contains a term
$d_{\pm}\equiv 1-\b n\b\beta(t\pm \tau/2)$. For small values of $\tau$, we have
$d_{\pm}= (1-\b n\b\beta)\mp \b n\dot{\b\beta\,}\frac{\tau}{2}$.
For ultrarelativistic particles the angle $\theta$ between $\b n$ and $\b\beta$
is small ($\sim1/\gamma$), thus, given the orthogonality of $\dot{\b\beta\,}$
and $\b \beta$, the last term in $d_{\pm}$ can be estimated as
$\frac{eB\lambda}{mc^2\gamma^2}$. Since this term was neglected at derivation of
equation~(\ref{jtt8}), it must be small compared to the first term in $d_\pm$,
which is estimated as $\sim 1/\gamma^2$. Thus, the range of applicability is
determined by the condition
\be\label{jtt10a}
\frac{eB\lambda}{mc^2}\ll 1 \, .
\ee

Furthermore, the exponential term in equation~(\ref{jitt2}) contains a
function, $i\Delta\equiv i(\Phi(t+\tau/2)-\Phi(t-\tau/2))$. The Taylor expansion
 of the function $\Delta$ gives $\Delta=\omega(1-\b n\b\beta)\tau-\omega\b
n\ddot{\b\beta\,}\tau^3/24$. In derivation of equation~(\ref{jtt4}) only the
first term in this expansion has been kept, therefore the applicability can be
reduced to the condition of neglecting the second term. Since the function
$\Delta$ is in the exponent, the condition is $\omega\b
n\ddot{\b\beta\,}(\lambda/c)^3\ll1$. The module of the particle velocity in
magnetic field remains constant, $\b\beta^2={\rm const}$, thus
$\frac12\,\frac{d^2}{dt^2}\b\beta^2=(\dot{\b\beta\,})^2+\b\beta\ddot{\b\beta\,}
=0$. Since emitted photons and the particle velocities are nearly parallel, $\b
n\approx \b\beta$, the term $(\b n\ddot{\b\beta\,})$ can be estimated as $(\b
n\ddot{\b\beta\,})\approx(\b\beta\ddot{\b\beta\,})=-(\dot{\b\beta\,})^2$. This
gives the second condition of applicability of equation~(\ref{jtt8}):
$\omega(\dot{\b\beta\,})^2(\lambda/c)^3\ll1$. By expressing the acceleration
$\dot{\b\beta\,}$ through the
magnetic field strength, the condition of applicability of
equation~(\ref{jtt8}) can be written in the form
\be\label{jtt10b}
\omega\ll \frac{m^2 c^5 \gamma^2}{e^2B^2\lambda^3}\,.
\ee
Note that for a homogeneous magnetic field $\lambda=\infty$, therefore the
standard synchrotron spectrum cannot be derived in the framework of
perturbation theory.

Equations (\ref{jtt10a}) and (\ref{jtt10b}) as conditions of applicability of
the perturbation approach can be interpreted in the following way. If a charged
particle travels in a region filled with magnetic field a path $\lambda$
which is shorter compared to the trajectory curvature $R$, then the particle
is deflected by an
angle $\d\theta \simeq \lambda/R$. The first condition given by
equation~(\ref{jtt10a}) implies that $\d\theta\ll 1/\gamma$.
Concerning the second condition given by equation~(\ref{jtt10b}), it is
equivalent to the requirement that the segment of the trajectory of length of
order
$\sim \lambda$, can be treated as a straight line.

The characteristic frequency of the radiation in this regime, $\omega_j$, can be
estimated from first principles. Namely,
while the emission is formed during the time interval $\d t_{\rm rad}\sim
\lambda/c$, it is registered during
$\d t_{\rm obs} =\d
t_{\rm rad}\,(1-\b n\b\beta)\sim \d t_{\rm rad}/\gamma^2$. Thus the
characteristic frequency is estimated
as \citep[see e.g.][]{Landau2}
\be\label{jitter1}
\omega_j=\frac{1}{\d t_{\rm
obs}}=\frac{c\gamma^2}{\lambda}\,.
\ee
Note that this frequency $\omega_j$ is independent on the magnetic field
strength $B$.

It is interesting to compare the characteristic frequencies, at which the bulk
of radiation is produced,
in highly turbulent and homogeneous magnetic fields corresponding to the jitter
and synchrotron
radiation regimes. The characteristic synchrotron frequency can be expressed
through the nonrelativistic Larmor radius $R_L=mc^2/eB$:
\be\label{jitter2}
\omega_c=\frac{3c\gamma^2}{2\,R_L}=\frac32\,\frac{\lambda}{R_L}\,\omega_j\, .
\ee

It is convenient to express equations~(\ref{jtt10a}) and (\ref{jtt10b}) also
through $R_L$:
\be\label{jitter3}
\frac{\lambda}{R_L}\ll 1\,,\qquad \omega\ll
\omega_j\left(\frac{R_L}{\lambda}\right)^{\!2}\,,
\ee
When these conditions are satisfied, the ratio $\omega_j/\omega_c\sim
R_L/\lambda \gg1$, i.e. the characteristic energy of photons emitted by charged
particles in highly turbulent magnetic field may significantly
exceed, by a factor of $R_L/\lambda$, the characteristic
energy of synchrotron photons emitted by same particles in a regular magnetic
field of same strength.

Finally, one should mention another constraint on applicability of
equation~(\ref{jtt8}) related to the plasma effects. The basic
equation~(\ref{jitt0}) describes emission in {\it vacuum}
neglecting the impact of the surrounding plasma.
If the radiating particle is located in plasma, the latter in the frequency
range $\omega \gg\omega_p$
can be treated as a medium with dielectric permittivity
$\e(\omega)=1-\omega_p^2/\omega^2$, where
\be
\omega_p=\sqrt{\frac{4\pi e^2n_e}{m_e}}\,,
\ee
is the plasma frequency ($n_e$, $m_e$ and $e$ are the number density, mass and
charge of electrons, respectively).
At $\e(\omega) \approx 1$, the term $(1-\b
n\b\beta)\approx(\frac1{\gamma^2}+\theta^2)/2$ in all above derived formulas
should be replaced by the one corrected for the dielectric permittivity,
$(1-\sqrt{\e}\b n\b\beta)\approx(\frac1{\gamma^2}+\frac{\omega_p^2}
{\omega^2}+\theta^2)/2$.
Thus, the influence of the medium can be ignored for sufficiently high
frequencies, $\omega\gg \omega_p\gamma$.
Note that the particle Lorentz factor $\gamma$ and the plasma frequency enter
the equation in the form of the
combination $1/\gamma^2+(\omega_p/\omega)^2$. Therefore, the influence of the
medium can be taken into account
if in all above equations we replace the particle Lorentz factor $\gamma$ to
\be\label{gamma_star}
\gamma^*(\omega)=\frac{\gamma}{\sqrt{1+(\omega_p\gamma/\omega)^2}}\,.
\ee
\citep[see e.g.][]{Ginz_Syr, Fleish06}.
However, as long as we are interested in high frequency range $\omega\gg
\omega_p\gamma$,
for the sake of simplicity we will ignore (unless otherwise is stated) the
difference between $\gamma$ and $\gamma^*$.

\section{Dealing with Chaotic Magnetic Field}
\label{sec:chaotic}
The integrand of equation~(\ref{jtt8}) contains the term
\[
(\b a_+\b a_-)=\frac{e^2}{m^2\gamma^2}\,
(\b\beta\times \b B_+)(\b\beta\times \b B_-)
\]
\be\label{jitter4}
=\frac{e^2\beta^2}{m^2\gamma^2}\,
(\d_{\rho\sigma}-\nu_\rho\nu_\sigma)\,B_{+\rho}B_{-\sigma}\,,
\ee
which should be averaged over different configurations of the magnetic field.
Here $\b\nu=\b\beta/|\b\beta|$ is unit velocity vector. The magnetic field
values $\b B_{+}$ and $\b B_{-}$ corresponds to the points where
the charged particle is located at time instants $(t\pm \tau/2)$, i.e. $\b
B_{\pm}=\b B(\b r(t)\pm \b\beta(t)\tau/2,t\pm \tau/2)$. The statistical
averaging of this expression will result in appearance of the {\it
correlation function}:
\be\label{jtt11}
K_{\rho\sigma} \equiv 
\langle B_{\rho}(\b r_1,t_1)B_{\sigma}(\b r_2,t_2)\rangle\,,
\ee
which is a second order tensor. Here, under the statistical averaging we
suppose a general standard procedure; it could
be a space-time homogenization or an integration over an ensemble of field
configurations \citep[see e.g. \S~118 of ][]{Landau5}. Here we assume that the
field is statistically homogeneous and stationary. This implies that the
correlation function depends only on the difference of the coordinates $(\b
r_1-\b r_2)$ and times $(t_1-t_2)$, i.e. $K_{\rho\sigma}=K_{\rho\sigma}(\b
r_1-\b r_2,t_1-t_2)$. In this case, $\langle\b B^2\rangle= K_{\rho\rho}(0)={\rm
const}$.

It is convenient to present the correlation function $K_{\rho\sigma}$ in the
form of a Fourier integral:
\be\label{jtt12}
K_{\rho\sigma}(\b r,t)= \int \t K_{\rho\sigma}(\b q,\k)\,
e^{i(\b q\b r-\k t)}\,\frac{d^3q}{(2\pi)^3}\,\frac{d\k}{2\pi}\,.
\ee
Since the magnetic field is divergence free ($\nabla\b B=0$), $K_{\rho\sigma}$
should satisfy the following conditions
\be\label{jtt13}
\p K_{\rho\sigma}/\p x_\rho=0\,,\qquad \p K_{\rho\sigma}/\p x_\sigma=0\,,
\ee
which for the Fourier transform $\t K_{\rho\sigma}$ take on form
({\it the transversality condition}):
\be\label{jtt13a}
\t K_{\rho\sigma}q_\rho=0\,,\qquad \t K_{\rho\sigma}q_\sigma=0\,.
\ee

While in Section~\ref{sec:prev} we will briefly discuss different tensor
structures of the correlation function,
here we consider the case of isotropic turbulence. This results in the
following
form of the correlation function \citep[see e.g.][]{Fleish06a}):
\be\label{jtt14}
\t K_{\rho\sigma}(\b q,\k)=\frac12\left(\d_{\rho\sigma}-
\frac{q_\rho q_\sigma}{q^2} \right) \Psi(|\b q|,\k)\langle\b B^2\rangle.
\ee
Here the constant factor $\langle\b B^2\rangle$ is introduced which allows
$\Psi$
to meet the normalization condition
\be\label{jtt15}
\int\! \Psi(q,\k)\,\frac{d^3q}{(2\pi)^3}\,\frac{d\k}{2\pi}=
\frac1{4\pi^3}\int\limits_{-\infty}^\infty\! d\k\int\limits_0^\infty\!
dq\,q^2 \Psi(q,\k)=1\,.
\ee
The tensor structure given by equation (\ref{jtt14}), obviously
satisfies the transversality condition of equation~(\ref{jtt13a}).

The averaged values of $(\b a_+\b a_-)$ can be expressed through the
correlation function as
\be\label{jtt16}
\langle\b a_+\b a_-\rangle = \frac{e^2}{m^2\gamma^2}\,
(\d_{\rho\sigma}-\nu_\rho\nu_\sigma)\,K_{\rho\sigma}(c\b\beta\tau,\tau)\,
.
\ee
Here we took into account that $\b r_+-\b r_-=c\b\beta\tau$, and replaced in
the numerator $\beta^2$ to 1. From equations~(\ref{jtt12})~and~(\ref{jtt14})
we find
\ba
&\ds\int\limits_{-\infty}^\infty \! \langle(\b a_+\b a_-)\rangle\,
 e^{i\t\omega \tau}\,d\tau =
\frac{e^2}{2m^2\gamma^2}\,\langle\b B^2\rangle &\nonumber\\
&\ds\hspace{-18pt}\!\times\!\!\!\int\!\! \left(1+
\frac{(\b\nu\b q)^2}{q^2}\right) \Psi(q,\k) e^{i(c\b
q\b\beta-\k+\t\omega)\tau}
\,\frac{d^3q}{(2\pi)^3}\,\frac{d\k}{2\pi} \,d\tau\,.&
\ea
After substitution of $\b\beta$ by the velocity unit vector $\b\nu$, and
integration over $d\tau$ resulting in a $\d$-function
$2\pi\,\d(c\b q\b\nu-\k+\t\omega)$, the integral over $d\k$ can be computed
analytically:
\[
\int\limits_{-\infty}^\infty \! \langle \b a_+\b a_-\rangle\,
 e^{i\t\omega \tau}\,d\tau =
\frac{e^2}{2m^2\gamma^2}\,\langle\b B^2\rangle
\]
\be\label{jtt17}
\times\int \! \left(1+\frac{(\b\nu\b q)^2}{q^2}\right) \Psi(q,\t\omega+c\b
q\b\nu)
\,\frac{d^3q}{(2\pi)^3}\,.
\ee
Thus, we arrive at the following expression for the energy and angular
distribution
of radiation per unit time
\ba
&\ds P_{\b n\omega}(t) = \frac{e^4}{2\pi^2m^2c^3}\,\langle \b B^2\rangle
\frac{\gamma^2(1+\gamma^4\theta^4)}{(1+\gamma^2\theta^2)^4}& \nonumber\\
&\ds \times\int \left(1+\frac{(\b\nu\b q)^2}{q^2}\right) \Psi(q,\t\omega+c\b
q\b\nu)
\,\frac{d^3q}{(2\pi)^3}\,, \label{jtt18} &
\ea
where $\t\omega$ is determined by equation~(\ref{jtt9}).

Let's consider now the case of steady turbulence, i.e., when the correlation
function given by equation~\eqref{jtt11}
is time-independent. Then the Fourier image of the correlation function
contains a $\d$-function,
$\Psi(q,\k)=\Psi(q)\,2\pi\d(\k)$, and the normalization condition (\ref{jtt15})
becomes
\be\label{jtt15a}
\int\! \Psi(q)\,\frac{d^3q}{(2\pi)^3}=
\frac1{2\pi^2}\int\limits_0^\infty\! \Psi(q)\,q^2\,dq=1\,.
\ee
Note that the function $\Psi$ determines the spectrum of the energy density of
the stochastic magnetic field, since
\be
\frac{\langle\b B^2\rangle}{8\,\pi}=
\frac{\langle\b B^2\rangle}{16\,\pi^3} \int\limits_0^\infty \Psi(q)\,q^2\,dq\,.
\ee
In the case of the stationary turbulence, a $\d$-functional factor, $2\pi\d(\t
\omega+c\b\nu\b q)$,
appears in the integrand of equation~(\ref{jtt18}). This makes the integration
over
$d\Omega_{\b q}$ (note that $d^3q=q^2dqd\Omega_{\b q}$) rather trivial:
 \[
P_{\b n\omega}(t)=
\frac{e^4}{4\pi^3 m^2c^4}\,\langle \b B^2\rangle
\frac{\gamma^2(1+\gamma^4\theta^4)}{(1+\gamma^2\theta^2)^4}
 \]
\be\label{jtt22}
\times\int\limits_{\t\omega/c}^\infty \!\left(1+\frac{\t\omega^2}
{c^2q^2}\right)\Psi(q) \,q\,dq\,.
\ee

Now we can conduct analytical integration over the emitting angles of
radiation. The major contribution to the integral comes from range of small
angles $\theta\lesssim 1/\gamma$; the contribution from large angles,
$\theta\gg 1/\gamma$, is negligibly small. Thus, applying the standard approach
for calculations of radiation of ultra-relativistic particles,
one can adopt $d\Omega=2\pi\theta\,d\theta$, and perform integration over
$\theta$ from zero to infinity.
It is also convenient to introduce a new integration variable
$\zeta=\gamma^2\theta^2$ and
change the order of integration over $\zeta$ and $q$. After performing a
trivial integration over $\zeta$,
we arrive at
\be\label{jtt23}
P_{\omega}(t)= \frac{e^4\langle \b B^2 \rangle}{6\pi^2 m^2c^4}
\int\limits_{\omega/(2c\gamma^2)}^\infty \! u(\xi)\, \Psi(q) \,q\,dq\,,
\ee
where $\xi=2qc\gamma^2/\omega$, and
\be\label{jtt23a}
u(\xi)=1+\frac3{\xi^2}-
\frac{4}{\xi^3}-\frac{3\ln\xi}{\xi^2}\,.
\ee

In the range of integration over $dq$, the variable $\xi$ alters from 1 to
$\infty$; while the function $u(\xi)$ increases monotonically from $u(1)=0$
to $u(\infty)=1$. Adopting $\xi$ as the integration variable,
equation~(\ref{jtt23})
can be presented in the form convenient for numerical computations:
\be\label{jtt25}
P_\omega(t) =\frac{e^4\langle \b B^2 \rangle}{6\pi^2 m^2c^4}\,
\int\limits_1^\infty \!u(\xi)\, \left(\frac{\omega\xi}{2c\gamma^2}\right)^{\!2}
\Psi\!\left(\frac{\omega\xi}{2c\gamma^2}\right) \frac{d\xi}{\xi}\, ,
\ee

Equation~(\ref{jtt23}) is an integral function which depends on the turbulence
spectrum.
However, it obeys some general properties not affected by the turbulence. In
particular, from equation~\eqref{jtt23}
follows that independent of $\Psi(q)$, the radiation spectrum $P_\omega(t)$ is a
 monotonically decreasing function of $\omega$. This feature becomes obvious
after the differentiation over $\omega$:
\be
\frac{\p P_\omega}{\p \omega}\,\sim \!
\int\limits_{\omega/(2c\gamma^2)}^\infty \!\!
\frac{\p\xi}{\p\omega}
\frac{du(\xi)}{d\xi}\, \Psi(q) \,q\,dq\, .
\ee
Here it is taken into account that the contribution to the derivative from the
lower integration limit is null (given that $u(1)=0$). Since $du(\xi)/d\xi>0$,
$\Psi(q)\ge 0$ and $\p\xi/\p\omega<0$, the integrand is negative, and the
integration
results in $\p P_\omega/\p\omega<0$. Thus, this function achieves its maximum
value at $\omega=0$, i.e.,
\be\label{jtt23aa}
P_{\omega}(t) \le P_{0} = \frac{e^4\langle \b B^2
\rangle}{6\pi^2 m^2c^4} \int\limits_{0}^\infty \! \Psi(q) \,q\,dq\,.
\ee
Of course, this estimate is meaningful only if the integral in the right side
of equation converges.

Note that the photon energy and the particle Lorentz factor enter to
equation~\eqref{jtt23}
only in the form of ratio $\omega/\gamma^2$. Thus, the spectrum $P_\omega$ is,
in fact, a function of one argument $\omega/\omega_j$ (for intermediate
calculations we drop, just for simplicity, the argument $t$):
\be\label{jtt24}
P_{\omega} \equiv \t P\!\left(\frac{\omega}{\omega_j}\right)=
\t P\!\left(\frac{\lambda\omega}{c\gamma^2}\right)\, .
\ee
Here $\t P(\omega/\omega_j)$ is a monotonically decreasing function.
In case of absence of other characteristic frequencies in the physical setup
except for $\omega_j$,
in the range $\omega\ll \omega_j$ function $\t P$ is nearly constant,
 $\t P\approx P_0$.
However, at very small frequencies
the surrounding plasma may significantly change the behavior of
$P_\omega$.
The substitution of $\gamma$ by $\gamma_*$ (in accordance with
equation~(\ref{gamma_star}))
leads to equation~(\ref{jtt23}) in which $\xi$ should be replaced by
$\xi_*=2qc\gamma_*^2/\omega$.

\begin{figure}
\centering{
\includegraphics[width=0.33\textwidth,angle=-90]{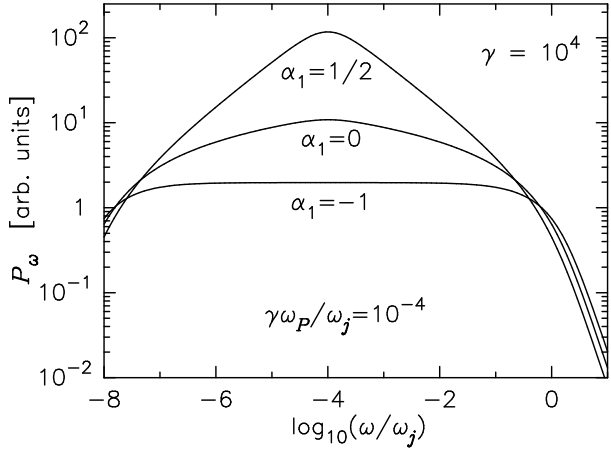}
}
\caption{\small Spectral power calculated for the turbulence spectrum $\Psi$
given in the form of equation~(\ref{nu1_0_1}) for three different values of
power-law
index $\alpha_1^{}$.
\label{fig:nu1_0_1}
}
\end{figure}

The derivative
\be
\frac{\p\xi_*}{\p\omega}=-2qc\gamma^2\,\frac{\omega^2-\gamma^2\omega_p^2}
{(\omega^2+\gamma^2\omega_p^2)^2}\,,
\ee
has a positive sign at $\omega<\gamma\omega_p$, and becomes negative
when $\omega>\gamma\omega_p$. Therefore,
independent of the choice of the spectrum of turbulence $\Psi(q)$,
the emission intensity increases with frequency
in the range of $\omega<\gamma\omega_p$, and decreases when
$\omega>\gamma\omega_p$, while the maximum is reached at
$\omega=\gamma\omega_p$.
Then, instead of equation~(\ref{jtt24}), we have
\be\label{jtt24a}
P_{\omega} =
\t P\!\left[\frac{\lambda}{c\gamma^2}
\left(\omega+\frac{\gamma^2\omega_P^2}{\omega}\right)\right].
\ee
The argument of this function has a minimum at $\omega = \gamma\omega_p$, and
consequently the function achieves its maximum at this frequency. However, we
should note that in the case of convergence of the integral in
equation~(\ref{jtt23a}), this maximum
would be practically invisible. To demonstrate the behavior of $P_\omega$ at
small frequencies,
in figure~\ref{fig:nu1_0_1} we show calculations
for three different turbulence spectra $\Psi$ presented in the following
specific form
\be\label{nu1_0_1}
\Psi(q)=\frac{A_{\alpha_1}}{q^{2+\alpha_1^{}}(1+\lambda^2q^2)^{1-\alpha_1^{}/2}}
\,.
\ee
Here, according to equation~(\ref{jtt15a}), the normalization constant
\be
A_{\alpha_1}= 4\pi^{3/2} \lambda^{1-\alpha_1} \,
\frac{\Gamma(1-\alpha_1/2)}{\Gamma((1-\alpha_1)/2)}\,,
\ee
where $\Gamma(z)$ is the gamma-function. The results of calculations in
figure~\ref{fig:nu1_0_1} correspond to
 three different values of $\alpha_1^{}$: $\alpha_1^{}=-1$, 0 and 1/2. It can be
seen that while
for $\alpha_1^{}=0$ or 1/2 the integral in equation~(\ref{jtt23a}) diverges and
the maximum of $P_\omega$ is clearly seen at $\gamma\omega_p$, for the value of
$\alpha_1^{}=-1$ the emission intensity is characterized by a broad plateau
without any distinct maximum.

To explore the emission spectra in the frequency range
$\gamma\omega_p\ll\omega\ll \omega_j$ and $\omega\gg\omega_p$, let's
assume that the turbulence spectrum has a broken power-law form:
\be\label{asympt_psi}
\Psi(q)=\left\{
\begin{array}{ll}
\lambda^3\left(\frac{q_1}{q}\right)^{\!2+\alpha_1}\,,&\qquad
q\ll\frac{1}{\lambda}\,,\\
\lambda^3\left(\frac{q_2}{q}\right)^{\!2+\alpha_2^{}}\,,&\qquad
q\gg\frac{1}{\lambda}\,,
\end{array}
\right.
\ee
where $q_1$ and $q_2$ are constants of the order of $1/\lambda$, and the factor
$\lambda^3$ is introduced for the reason of
dimension consistency. The condition for the
convergence of the integral in equation~(\ref{jtt15a}) on the lower and upper
limits implies
\be
\alpha_1^{}<1\,,\qquad \alpha_2^{}>1\,.
\ee

Depending on the value of $\alpha_1$ there are two different cases related to
the convergence of the integral in equation~\eqref{jtt23a}. If the integral is
converging at the lower limit (i.e., $\alpha_1^{}<0$), we have the case
discussed above. Let's consider now the range of $0<\alpha_1^{}<1$. Then, for
the frequency interval $\gamma\omega_P^{}\ll\omega\ll\omega_j$
we have
\be\label{jittas1}
P_\omega =
\frac{e^4\lambda^3q_1^2\langle\b B^2\rangle}{2\pi^2 m^2c^4}\,
\left[\left(\frac{2c\gamma^2q_1}{\omega}\right)^{\!\alpha_1}
-1\right] \frac{C_1}{\alpha_1^{}}\,,
\ee
where
\be
C_1=\frac{4+3\alpha_1^{}+\alpha_1^2}{(3+\alpha_1^{})(2+\alpha_1^{})^2}\,.
\ee
At lower frequencies, $\omega_P^{} \ll\omega\ll \gamma\omega_P^{}$,
\be\label{jittas2}
P_\omega =
\frac{e^4\lambda^3q_1^2\langle\b B^2\rangle}{2\pi^2 m^2c^4}\,
\left[\left(\frac{2c \omega q_1}{\omega_P^2}\right)^{\!\alpha_1}-1\right]
\frac{ C_1}{\alpha_1^{}}\,.
\ee
We note that these equations~(\ref{jittas1}, \ref{jittas2}) allow a smooth
passage to the limit $\alpha_1^{}\to 0$.

In the range of large frequencies, $\omega\gg \omega_j$,
the radiation spectrum behaves as a power-law $\omega^{-\alpha_2}$,
i.e. mimics the turbulence spectrum \citepalias{Toptygin_Fleish},
\be
P_\omega =
\frac{e^4\lambda^3q_2^2\langle\b B^2\rangle}{6\pi^2 m^2c^4}\,
\left(\frac{2c\gamma^2q_2}{\omega}\right)^{\!\alpha_2} C_2\,,
\ee
where
\be
C_2=\frac1{\alpha_2}+\frac3{1+\alpha_2}-\frac3{(2+\alpha_2)^2}-\frac4{3+\alpha_2
}\,.
\ee

The energy lose rate of a charged particle due to radiation in the magnetic
field is
given by the classical formula \cite[see e.g.][]{Landau2}
\be\label{jtt26}
I=\frac{2e^4\gamma^2}{3m^2c^3}\,(\b\beta\times \b B)^2=
\frac{2e^4\beta^2\gamma^2}{3m^2c^3}\, B^2\sin^2\chi\,,
\ee
where $\chi$ is angle between the particle velocity and direction of the
magnetic field.
By averaging $I$, first over the directions, then over the strength of the
magnetic field, and taking into
account that $\langle\sin^2\chi\rangle=\frac23$, one finds
\be\label{jtt27}
I=\frac{4e^4\gamma^2}{9m^2c^3}\, \langle B^2\rangle\,
\ee
(in the numerator, $\beta^2=1$ is substituted). By definition, the same
result can be obtained by direct integration
of equation~(\ref{jtt23}) or equation~(\ref{jtt25}) over the emitted photon
frequencies: $\int_0^\infty\!P_\omega\,d\omega=I$. Nevertheless, it is worth to
perform we such computations; they can serve as a good test for
the consistency of the results.

From equation~(\ref{jtt25}) we find
\be\label{jtt28}
\int\limits_0^\infty\! P_\omega\,d\omega =\frac{e^4\langle \b B^2\rangle}
{6\pi^2 m^2c^4}\!\int\limits_1^\infty \!\frac{d\xi}{\xi}\,u(\xi)\!
\int\limits_0^\infty\! d\omega\left(\frac{\omega\xi}{2c\gamma^2}\right)^{\!\!2}
\!\Psi\!\left(\frac{\omega\xi}{2c\gamma^2}\right) \!.
\ee
After the substitution of the new variable $\omega=q\times(2c\gamma^2)/\xi$,
the integration over $\omega$ leads to equation~(\ref{jtt15a}), and then we
obtain
\be\label{jtt29}
\int\limits_0^\infty\! P_\omega\,d\omega =
\frac{2e^4\gamma^2\langle \b B^2\rangle}
{3 m^2c^3} \int\limits_1^\infty \!\frac{d\xi}{\xi^2}\,u(\xi)\,.
\ee
The remaining integral is equal to $2/3$, so the direct integration of the
emission spectrum leads to equation~(\ref{jtt27}). This interesting result,
when the integration of the {\it approximate} equation~(\ref{jtt25}) gives
{\it precise} expression for energy losses, has a quite natural explanation;
while equation~(\ref{jtt25}) contains the first
(quadratic) term in the expansion of the spectrum over the magnetic field
strength $B$, the precise expression for the energy losses given by
equation~(\ref{jtt27}) is proportional to the second power of $B$. Also we note
 that the energy losses
are independent of the spectrum of turbulence $\Psi$.

In a similar way one can find the angular distribution of the emission after
integration over the frequencies; therefore
we simply write down the final result:
\be\label{jtt30}
dI_{\b n}= \frac{4e^4\gamma^4 \langle \b B^2\rangle}{3\pi m^2c^3}\,
\frac{1+\gamma^4\theta^4}{(1+\gamma^2\theta^2)^5}\, d\Omega\,.
\ee
Here, $dI_{\b n}$ is the energy emitted into the solid angle $d\Omega$ per
time unit.
This angular distribution also does not depend on the turbulence spectrum
$\Psi$.

\section{Large scale turbulence}
\label{sec:synch}

In the case of a large scale magnetic field turbulence, $\lambda\gtrsim R_L$,
the conditions imposed by equation~\eqref{jitter3} are violated, therefore
the results of the previous section are not anymore valid. On the other hand,
the radiation spectrum formed in the regime $\lambda\gg R_L$ can be
derived analytically.
In this case the particle deflection angle exceeds $1/\gamma$, and
the radiation spectrum, $P_\omega$, is determined by the instant curvature of
trajectory (or the instant value of the magnetic field).
Thus the result should be similar to the spectrum of synchrotron radiation
in the homogeneous magnetic field \citep{Schwinger, Ginz_Syr, Landau2}.
If the charged particle moves perpendicularly to the magnetic field, the
emission spectrum is determined as
\be\label{sy_ra22}
P_\omega(t)=\frac{\sqrt{3}\,e^2}{2\pi R_L} \,F(x)\,,
\ee
where
\begin{equation}\label{sy_ra24}
 F(x)=x\int\limits_x^\infty \!K_{5/3}(u)\,du\,.
\end{equation}
Here $K_{5/3}(u)$ is the modified Bessel function, $x=\omega/\omega_c$ and
$\omega_c$ is determined by equation~(\ref{jitter2}). If the charged particle
moves at an angle $\chi$ to the magnetic field,
in equation~({\ref{sy_ra22}) $B$ should be substituted by the
perpendicular component of the field, $B_\perp\equiv B\sin\chi$
\citep{Ginz_Syr}.

If the magnetic field is turbulent, then the spectrum $P_\omega(t)$ should be
averaged over directions of the field, i.e., integrated over the pitch angle
$\chi$. This results in the following expression \citep{Schlickeiser}
\be\label{sy_ra22a}
P_\omega(t)=\frac{\sqrt{3}\,e^2}{2\pi R_L} \,G(x)\,,
\ee
where
\be\label{sy_un6a}
G(x)=\frac{\pi x}{2}\left(W_{0,\frac43}(x)\,W_{0,\frac13}(x)-
W_{\frac12,\frac56}(x)\,W_{-\frac12,\frac56}(x)\right) \, .
\ee
Here $W_{\mu,\alpha}(x)$ is the Whittaker function.

The function $G(x)$ can be presented in a more convenient form:
\be\label{sy_un6}
G(x)=\frac{x}{20}\left[(8+3x^2)\,(\kappa_{1/3})^2+
x\,\kappa_{2/3}\,(2\kappa_{1/3}-3x\kappa_{2/3}) \right] \, ,
\ee
via familiar Bessel functions $\kappa_{1/3}=K_{1/3}(x/2)$,
$\kappa_{2/3}=K_{2/3}(x/2)$ \citep{AKP10}.
$G(x)$ has a simple asymptotic behavior both at low and high frequencies:
\be\label{un6x}
G(x)\approx \left\{
 \begin{array}{lll}
\frac{2^{1/3}}{5}\,\big(\Gamma(1/3)\big)^2x^{1/3}\,,&\quad x\ll 1, \\ [4pt]
\frac\pi2\,e^{-x}\,, &\quad x\gg 1. \\
 \end{array}
\right.
\ee

Although differences between the spectra of synchrotron radiation in
homogeneous and (large scale)
chaotic fields, i.e.
between functions $F(x)$ and $G(x)$, are not dramatic, yet they not too
small to be neglected in
calculations \citep{AKP10}. In particular, these functions achieve their
maximums,
$\max(F)=0.918$ and $\max(G)=0.713$, at different points, $x=0.286$ and
$x=0.229$, respectively.
Obviously, similar differences we should expect for the spectral energy
distributions described by the functions
$x\,F(x)$ and $x\,G(x)$. Namely, $\max(x\,F)=0.693$ is achieved at $x=1.33$,
and $\max(x\,G)=0.444$ is achieved
at $x=1.15$.

Finally, we note that function $G(x)$ can be approximated by a simple
analytical expression,
\be\label{sy_un11}
G(x)=\frac{1.808\,x^{1/3}}{\sqrt{ 1+3.4\,x^{2/3}}}\,
\frac{1+2.21\,x^{2/3}+0.347\,x^{4/3}}{1+ 1.353\,x^{2/3}+0.217\,x^{4/3}}\,
e^{-x}\, ,
\ee
which provides better than 0.2 \% accuracy \citep{AKP10}. Thus,
 this approximation can be safely used, instead of the precise
equation~(\ref{sy_un6}), in detailed calculations of radiation in environments
with large scale turbulent magnetic field.


When deriving Eq.(\ref{sy_ra22a}) we assumed that the magnetic field $\b B$
is oriented chaotically, but its absolute value, $|\b B|$, is fixed. However,
in a turbulent medium the spatial variation of the field
strength could be quite significant; therefore we have to average the results
also over the
absolute value of the field \cite[see, e.g.,][]{bykov2012}. Let's introduce the distribution
function
\be\label{diff_b1}
w(B)\,dB =h_n(b)\,dB/B_0\,,
\ee
where $B_0\equiv \sqrt{\langle \b B^2\rangle}$, $b=B/B_0$. By definition,
$w(B)\,dB$ is the probability of the strength of the magnetic field being in
the interval $(B,B+dB)$. We will consider three different distributions with
the function $h_n(b)$ presented in the following forms:
\be\label{diff_b2}
h_0(b)=\d(b-1)\,,
\ee
\be\label{diff_b3}
h_1(b)=\frac{3\sqrt{6}}{\sqrt{\pi}}\, b^2 e^{-3b^2/2}\,,
\ee
\be\label{diff_b4}
h_2(b)=\frac{32\,b^2}{\pi(1+b^2)^4}\,,
\ee
For all distributions
\be\label{diff_b5}
\int\limits_0^\infty\! h_n(b)\,db=\int\limits_0^\infty\! b^2 h_n(b)\,db=1\,.
\ee
For the variance of these distributions, $D=\langle b^2\rangle-\langle
b\rangle^2=
1-\langle b\rangle^2$, we have
\be\label{diff_b6}
D_0=0,\;\;D_1=0.15,\;\; D_2=0.28. 
\ee

The energy lose rate of particles given by Eq.(\ref{jtt27}) depends only on
$\langle \b B^2\rangle$,
therefore it is convenient to compare the averaged spectra for the same value of
$\langle \b B^2\rangle$:
\be\label{diff_b7}
\langle P_\omega\rangle\equiv \int\limits_0^\infty
P_\omega\,w(B)\,dB=I\,R_h(x)/\omega_0\,,
\ee
where $\omega_0=3eB_0\gamma^2/(2mc)$, $x=\omega/\omega_0$,
$\int_0^\infty\!R_h(x)\, dx=1$.

In Fig~\ref{fig:diff_b3} we show the SED of synchrotron radiation, $xR_h(x)$,
calculated for magnetic field distributions given by
Eqs.(\ref{diff_b2}),(\ref{diff_b3}) and (\ref{diff_b4}).

\begin{figure}
\centering{
\includegraphics[width=0.36\textwidth,angle=-90]{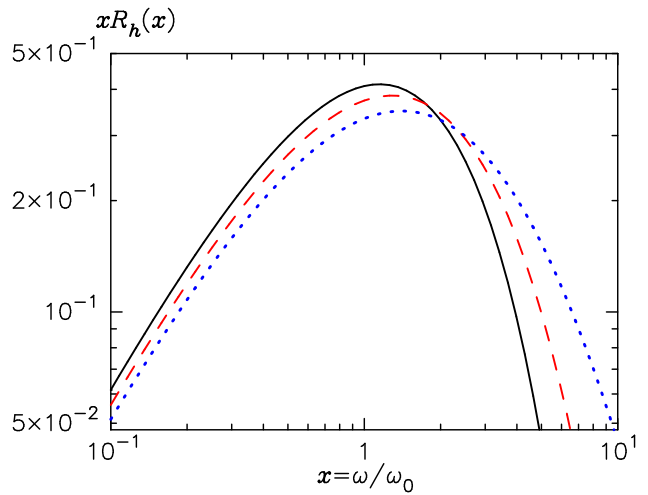}
}
\caption{Spectral energy distribution of synchrotron radiation corresponding to
distributions of the magnetic field $h_0$ (solid curve), $h_1$ (dashed curve),
$h_2$ (dotted curve).
}
\label{fig:diff_b3}
\end{figure}
One can see that the
spectrum of synchrotron radiation is somewhat shifted,
depending on the distribution of the magnetic field, towards higher energies,
and, more
importantly, it is significantly broadened compared to the radiation spectrum
relevant
to the $\delta$-functional distribution of magnetic field $h_0$.
One can show (see Appendix~\ref{app_a}) that if $w(B)$ is characterized by a power-lay asymptotic dependence: $w(B)\propto B^{-\sigma}$  for $B\rightarrow\infty$, then
the spectrum of synchrotron radiation also has a power-law asymptotic,
 namely $\langle P\rangle\propto \omega^{-\sigma+2}$  for $\omega\gg \omega_0$.
 Note that the power-law index should exceed $\sigma>3$ in order to provide
convergence of $\langle\b B^2\rangle$.

It is important to emphasize the broadening and  the shift of the spectrum of synchrotron radiation in a  
large-scale turbulent field has a quite different origin and should not be confused with the effects related to 
the jitter regime of radiation in a small-scale  turbulent field. In this regard we should note that a broadening of synchrotron 
radiation has been "observed" in the numerical simulations of  \citet{teraki_takahara_11}. However, most likely 
the authors   misinterpreted the obtained spectral feature and refereed it to an intermediate regime between 
the synchrotron and jitter regimes of radiation. However, we believe that 
this component of  radiation revealed in their simulations,  has a standard synchrotron origin, but 
simply broadened because of the distribution of the strength of the magnetic field.

\section{Energy spectra of radiation in the Jitter and Synchrotron regimes}
\label{sec:comp}
In this section we compare spectra of synchrotron and jitter radiation,
produced in two different
large- and small- scale turbulent magnetic fields but with the same
value $\langle \b B^2\rangle$, thus the total radiation power given by
equation~(\ref{jtt27})
is the same for radiation in both regimes. For comparison, it is convenient to
introduce the
normalized the emission intensity:
\be\label{sy_j_c1}
R(x)\,dx=P_\omega\,d\omega/I\,,\qquad x=\omega/\omega_c\,.
\ee
Obviously, the following condition is held: $\int_0^\infty\! R(x)\,dx=1$.

For calculations we  have to select a assume a  spectrum of turbulence and a distribution of the
magnetic field strengths. For the sake of  simplicity, below we consider
the case of chaotic synchrotron emission, i.e., we adopt a field distribution corresponding to 
equation (\ref{diff_b2}). Then the function $R$ depends only on the magnetic field,
\be\label{sy_j_c2}
R(x)=\frac{27\sqrt{3}}{16\pi}\,G(x)\,.
\ee
We note however that  $R$ as a function of argument $x$ does not depend on $\langle {\bf B^2}\rangle$.
In this section, for calculations of jitter radiation we consider  the turbulence spectrum as an
one-parameter family of functions:
\be \label{psi_q}
\Psi(q)=\frac{\lambda^3 A_\alpha}{(1+\lambda^2 q^2)^{1+\alpha/2}}\,.
\ee
The normalization constant, $A_\alpha$, is obtained from
equation~(\ref{jtt15a}):
\be \label{psi_qa}
A_\alpha=\frac{8\pi^{3/2}\,\Gamma(1+\alpha/2)}{\Gamma\big((\alpha-1)/2\big)}\,.
\ee
The spectrum presented in the form of equation~(\ref{psi_q}) is
characterized by a power-law dependence for $q\gg\lambda^{-1}$. Although
the spectra of turbulence, which can be generated in astrophysical
environments, remains an open question, usually it is approximated as
a power-law. This assumption is justified by a few
fundamental considerations. In particular, the power-law spectra of turbulence
with spectral indices of $5/3$ and $3/2$ appear in the
hydrodynamical \citep{kolmogorov41} and
magnetohydrodynamical \citep{iroshnikov63,kraichnan65} turbulent media.

Note that the asymptotic form of equation~(\ref{psi_q}) is consistent with
equation~(\ref{asympt_psi}) for
$\alpha_1^{}=-2$, $\alpha_2^{}=\alpha$. In figure~\ref{fig:jitt} we show the
normalized spectral energy distributions of the synchrotron and jitter
radiation, $x R(x)$
produced by particles of fixed energy $\gamma mc^2$. The spectra are plotted as
a function of $x=\e/\e_c=\hbar\omega/\hbar\omega_c$, for three different indices
characterizing the turbulence, $\alpha=2, 5/3, 3/2$.

For a rather broad range of variation of the index $\alpha$, from 3/2 to 3,
the presentation of the turbulence spectrum in the form of
equation~(\ref{psi_q}) allows simple analytical approximations for the
radiation power
\be \label{psi_q1}
P_\omega\,d\omega=I\,f(x_j)\,d\omega/\omega_j\,,
\ee
where $x_j=\omega/\omega_j$, and
\be \label{psi_q2}
f(x_j)=C_\alpha\,\big(1+0.22 \,x_j+0.43 \,x_j^2\big)^{-\alpha/2}\,.
\ee
The coefficient $C_\alpha$ is determined from the normalization $\int_0^\infty
f(x_j)\,dx_j=1$. The comparison with the
exact numerical calculations shows that the precision of this approximation is
better than 7\%.

\begin{figure}
\centering{
\includegraphics[width=0.43\textwidth,angle=0]{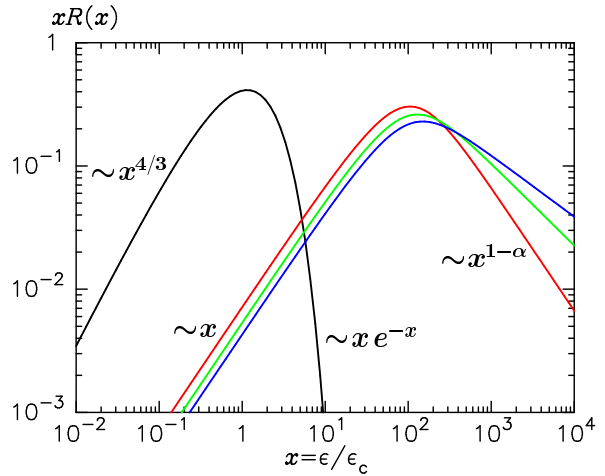} 
}
\caption{\small SED ($x R(x)$, see equation~\eqref{sy_j_c1}) of radiation of
monoenergetic particles
in turbulent magnetic field in the synchrotron and
jitter regimes: synchrotron (black line) and jitter (red, green and blue
solid lines).
The spectrum of turbulence was taken in the form of equation~(\ref{psi_q}). The
ratio of the correlation length to nonrelativistic Larmor radius ($R_L=mc^2/eB$) was adopted to be
$\lambda/R_L=10^{-2}$.
Red, green and blue lines correspond to the indices $\alpha=2$, 5/3
and 3/2, respectively.
}
\label{fig:jitt}
\end{figure}

Figure~\ref{fig:jitt} demonstrates the basic spectral features of the jitter
radiation. The SED peaks at energy which compared to the maximum of the
synchrotron radiation at $\epsilon=1.155 \epsilon_c$ is shifted by the factor
of $2/3\,R_L/\lambda$. Below the maximum $xR(x) \propto x$, i.e. the SED
increases
with energy slower than the SED of the synchrotron radiation, $xR(x) \propto
x^{4/3}$. Moreover, while the synchrotron spectrum has a quite sharp
(exponential) cutoff beyond $x \sim 1$, the SED of jitter radiation after the
break at $x \sim R_L/\lambda$ continues as a power-law, $x R(x) \propto
x^{1-\alpha}$ up to $x \sim (R_L/\lambda)^3$.

In astrophysical environments, acceleration of particles typically leads to
broad energy distributions. Below we compare the synchrotron and jitter
radiations for different distributions of accelerated particles $N(\gamma)$:
\be\label{sy_j_c3}
P(\omega)=\int\limits_0^\infty\!P_\omega\,N(\gamma)\,d\gamma\,.
\ee
Here we assume that energy distribution of all particles occupies certain energy
interval $(\gamma_{min},\gamma_{max})$. Outside this interval, the function $N$
is
null \footnote{We would like to indicate to the non-physical lower limit in the
integral in equation~\eqref{sy_j_c3}. However, this convenient for integration
representation is correct as long as the function $N$ is taken zero outside the
physically meaningful region}.
For the jitter radiation, using equation~(\ref{jtt25}) and introducing a new
dimensionless function $\Psi_1(\lambda q)=\Psi(q)/\lambda^3$, as well as
substituting the integration variable $\gamma$ by $\eta=\lambda\omega
\xi/(2c\gamma^2)$, we obtain
\[
P(\omega)=\frac{e^4\lambda\langle\b B^2 \rangle}{12\,\pi^2m^2c^4}
\!\int\limits_1^\infty\!d\xi\,u(\xi)
\]
\be\label{sy_j_c4}
\times\int\limits_0^\infty\! d\eta\,
\sqrt{\frac{\lambda\omega\eta}{2c\xi}}\Psi_1(\eta)\,
N\!\left(\sqrt{\frac{\lambda\omega\xi}{2c\eta}} \right).
\ee

Let's assume now that the relativistic charged particles have a power-law
distribution,
$N(\gamma)=N_0\gamma^{-\mu}$.
It can be shown that for the range of the power-law index,
$1<\mu<2\alpha+1$,
the main contribution to equation~(\ref{sy_j_c3}) is provided by particles of
energy
$\gamma\sim (\lambda\omega/c)^{1/2}$. Therefore, for the energy interval
$\omega\gg c/\lambda$, equation~(\ref{sy_j_c4}) can be
integrated over $d\eta$ in the limits from $0$ to $\infty$:
\[
P(\omega)=\frac{e^4\lambda\langle\b B^2 \rangle}{12\,\pi^2m^2c^4}
\left(\frac{2\,c}{\lambda\omega} \right)^{\!(\mu-1)/2}
\]
\be\label{sy_j_c5}
\times\!\int\limits_1^\infty\!d\xi\,u(\xi)\,\xi^{-(\mu+1)/2}
\int\limits_0^\infty\! d\eta\,\Psi_1(\eta)\,\eta^{(\mu+1)/2}\,.
\ee
The power-law dependence of the spectra ($P(\omega)\sim\omega^{-(\mu-1)/2}$) is
explained by the same reason as in the case of the synchrotron radiation:
$\omega$ and $\gamma$ enter into $P_\omega$ in a combined
form $\omega/\gamma^2$ \citep[for discussion of the case of synchrotron
radiation see][]{Rybicki}. Thus,
for a power-law particle distribution the synchrotron and jitter mechanisms lead
to
the same type of energy spectra, therefore the ratio of the emission
intensities due to these two processes,
\be
r\equiv\frac{P_{\rm jitt}(\omega)}{P_{\rm synchr}(\omega)} =
C(\mu,\alpha) \left(\frac{R_L}{\lambda} \right)^{\!(\mu-3)/2} \, ,
\ee
does not depend on photon energy $\omega$. Interestingly, the index of $\mu=3$
appears to be special, in the sense that independently of the turbulence
spectrum, the ratio $r=1$. This, in particular, can be seen
in figure~\ref{fig:jitt1_4} at low energies. Note that although the energy
losses
due to the synchrotron and jitter mechanisms
in the large and small turbulent fields are equal (for the same mean magnetic
field),
formally for $\mu>3$ larger
energy is radiated out due to the jitter mechanism ($r > 1$), and vise versa,
$r<1$ for $\mu<3$.
This apparent inconsistency is related to the assumption of pure power-law
particle distribution. However,
for a realistic distribution of particles with a high energy cutoff, the
spectral shape of the
synchrotron and jitter radiations differ significantly. In particular,
for power-law distributions with an exponential cutoff, given in a rather
general form
\be\label{pl_cutoff}
N\sim\gamma^{-\mu} \exp\big(-(\gamma/\gamma_{\rm cut})^\beta\big) \, ,
\ee
in the high energy limit, the shapes of the synchrotron and jitter radiations
spectra differ significantly (see figure~\ref{fig:jitt1_4}). While the
synchrotron
component beyond the maximum decreases exponentially \citep[][also see
\citealt{fritz89,za07}]{Lefa}:
\be\label{sy_j_c6}
P(\omega)\propto \exp\left[-\frac{\beta+2}{2}
\left(\frac{2\omega}{\omega_{\rm cut}}\right)^{\!\beta/(\beta+2)} \right],
\ee
with
\be\label{sy_cutoff}
\omega_{\rm cut}=\frac{3eB}{2mc}\gamma_{\rm cut}^2 \, ,
\ee
the jitter emission spectrum beyond the break around $\omega_{\rm cut}
(R_L/\lambda)$, has a long power-law tail, $P(\omega)\propto \omega^{-\alpha}$,
independently of the
shape of the particle distribution in the cutoff region (i.e. the value of
$\beta$). In this regard,
this is a unique feature of the jitter radiation which provides direct and
model-independent
information about the spectrum of turbulence. As long as the condition of
small-scale turbulence is satisfied
($\lambda < R_L$), we should expect radiation with characteristic a
broken power-law type spectrum. While the photon index
at low energies is directly related to the spectral index of relativistic
particles, $\Gamma=(\mu
 + 1)/2$, or in the case of a low energy cutoff or very hard particle spectrum
below the cutoff energy (e.g.
 in the case of Maxwellian type distribution - see figure~\ref{fig:jitt_vs_syn})
, $\Gamma=1$,
 the spectrum after the break depend only on the spectrum of turbulence. In an
environment with large scale
 turbulence, the picture is just opposite. The radiation proceeds in the
synchrotron regime and therefore
 is not sensitive to the details of the turbulence. On the other hand, the
synchrotron radiation carries information about the overall spectrum of parent
particles,
including the most important (from the point of view of the acceleration theory)
cutoff region.

\begin{figure}
\centering{
\includegraphics[width=0.33\textwidth,angle=-90]{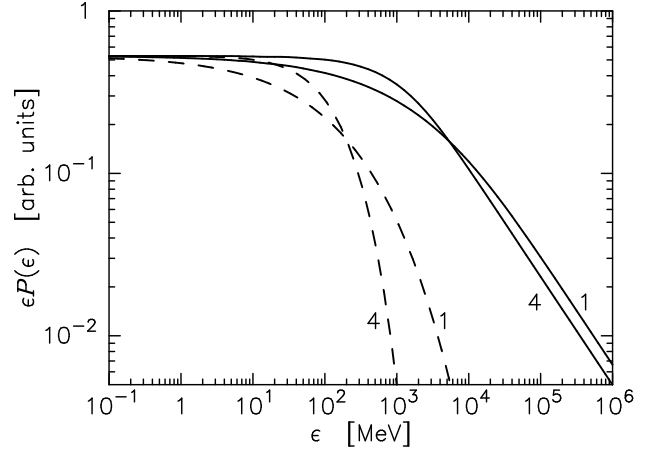}
}
\caption{\small SED of synchrotron (dashed lines) and jitter (solid lines)
radiation calculated for electron distribution:
$\gamma^{-3}\exp\!\left[-(\gamma/\gamma_{\rm cut})^\beta\right]$. Calculations
are performed for two values of $\beta$: $\beta=1$ and $\beta=4$
(the number labels indicate the used values for different lines ). The cutoff
energy is set $\gamma_{\rm cut}=10^8$, and the computations are
performed for $B=1$~G. The jitter radiation is computed for turbulence
spectrum in the form of equation~\eqref{nu1_0_1} with $\alpha=5/3$ and the ratio
of the
field correlation length to the nonrelativistic Larmor radius ($R_L=mc^2/eB$) is $\lambda/R_L=0.1$.}
\label{fig:jitt1_4}
\end{figure}

\begin{figure}
\centering{
\includegraphics[width=0.33\textwidth,angle=-90]{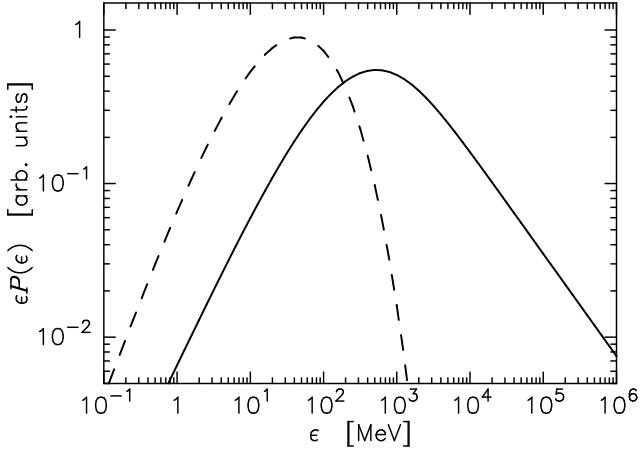}
}
\caption{\small The same as in figure~\ref{fig:jitt1_4}, but for Maxwellian-type
distribution of charged particles:
$\gamma^2\exp\!\left(-\gamma/\gamma_{\rm cut}\right)$ with
 $\gamma_{\rm cut}=10^7$.}
\label{fig:jitt_vs_syn}
\end{figure}

In Fig.~\ref{fig:jitt_vs_syn} we show a comparison of the synchrotron and jitter
radiation for the case of Maxwell distribution of emitting particles.

\section{Anisotropic Turbulence}\label{isotrop}

If the distribution of the charged particles is isotropic, the analytical
solutions derived in the previous sections
can be generalized to the case of the correlation tensor with an arbitrary
angular structure.
Indeed, similarly to equation~(\ref{jtt18}), the radiation power can be
expressed as
\ba
&\ds P_{\b n\omega}(t) = \frac{e^4}{\pi^2m^2c^3}\,
\frac{\gamma^2(1+\gamma^4\theta^4)}{(1+\gamma^2\theta^2)^4}\,(\d_{\rho\sigma}
-\nu_\rho\nu_\sigma)& \nonumber\\
&\ds \times\int \t K_{\rho\sigma}(\b q,\k) \,2\pi\,\d(\t\omega+c\b q\b\nu-\k)
\,\frac{d^3q}{(2\pi)^3}\,\frac{d\k}{2\pi}\,. \label{iso1} &
\ea
To obtain the radiation spectrum, this equation should be integrated over the
photon emitting angles and
averaged over directions of velocities of emitting particles, $\b\nu\equiv
\b\beta/|\b\beta|$. In order to simplify
calculations, let's introduce the following intermediary tensor
\ba
&\ds T_{\rho\sigma}=\frac12\! \int\!
(\d_{\rho\sigma}-\nu_\rho\nu_\sigma)\,
\frac{\gamma^2(1+\gamma^4\theta^4)}{(1+\gamma^2\theta^2)^4}&\nonumber\\
&\ds\times\d(\t\omega+c\b q\b\nu-\k)\,d\Omega\,d\Omega_\alpha\,,\label{iso2}&
\ea
where $d\Omega$ and $d\Omega_\alpha$ are the solid angles related to the
directions of momenta of the emitted photon
and the emitting particle, respectively. Note that equation~(\ref{iso2})
contains all the ``directional'' terms.
Thus, the radiation spectral power can be expressed as
\be\label{iso3}
P_{\omega}(t) = \frac{e^4}{\pi^2m^2c^3}\,
\int T_{\rho\sigma}\, \t K_{\rho\sigma}(\b q,\k)
 \,\frac{d^3q}{(2\pi)^3}\,\frac{d\k}{2\pi}\,.
\ee
 The correlation tensor $K_{\rho\sigma}$ and its Fourier transformation $\t
K_{\rho\sigma}$ are defined in Sec. 3.

According to equation~(\ref{iso2}), the tensor $T_{\rho\sigma}$ is a {\it
symmetric tensor} of the second order;
it depends only on the vector $\b q$, therefore, $T_{\rho\sigma}$ has the
following structure:
\be\label{iso4}
T_{\rho\sigma}=F_1\d_{\rho\sigma}+F_2q_\rho q_\sigma\,,
\ee
where $F_1$ and $F_2$ are scalar functions. The convolution of tensors
$T_{\rho\sigma}$ and
$\t K_{\rho\sigma}$, taking into account the transversality condition of
equation~(\ref{jtt13a}), gives
\be\label{iso5}
T_{\rho\sigma} \t K_{\rho\sigma}=F_1\t K_{\rho\rho}\,,
\ee
This expression determine the integrand in equation~(\ref{iso3}).

To obtain the scalar function $F_1$ the following relations can be used:
\be\label{iso6}
T_{\rho\rho}=3F_1+q^2F_2\,,\quad q_\rho q_\sigma T_{\rho\sigma}
=q^2F_1+q^4F_2\, ,
\ee
which give
\be\label{iso6a}
F_1=\frac12\left(T_{\rho\rho}- q_\rho q_\sigma T_{\rho\sigma}/q^2\right) \, .
\ee
Using equation~(\ref{iso2}), one finds
\[
F_1=\frac14\! \int\! \left(1+\frac{(\b q\b\nu)^2}{q^2} \right)
\]
\be\label{iso7}
\times
\frac{\gamma^2(1+\gamma^4\theta^4)}{(1+\gamma^2\theta^2)^4}
\,\d(\t\omega+c\b q\b\nu-\k)\,d\Omega\,d\Omega_\alpha\,,
\ee
which after the integration can be presented in the form (see Appendix~\ref{app_b})
\be\label{iso8}
F_1=\frac{\pi^2}{3qc}\,U\big(\xi,\k/(qc)\big)\,.
\ee
Here $U$ is determined by equation~(\ref{app4}).

The trace of the correlation tensor can be represented as
\be\label{iso9}
\t K_{\rho\rho}(\b q,\k)=\langle \b B^2\rangle\,\Psi(\b q,\k)\,,
\ee
where $\Psi(\b q,\k)$ satisfies the normalization condition:
\be
\int\! \Psi(\b q,\k)\frac{d^3q}{(2\pi)^3}\,\frac{d\k}{2\pi}=1\,.
\ee
Thus, the radiation power can be represented as
\be\label{iso10}
P_{\omega}(t)= \frac{e^4\langle \b B^2 \rangle}{3 m^2c^4}
\int\!\frac{1}{q}\,U(\xi,\k/qc)\, \Psi(\b q,\k)\, \frac{d^3q}{(2\pi)^3}
\,\frac{d\k}{2\pi}\,.
\ee

If the correlation function $K_{\rho\sigma}(\b r,t)$ does not depend on time,
i.e.
$\Psi(\b q,\k)=2\pi\d(\k)\,\Psi(\b q)$, then the integration over $\k$ is
trivial leading to
\be\label{iso11}
P_{\omega}(t)= \frac{e^4\langle \b B^2 \rangle}{24\pi^3 m^2c^4}
\int \! u(\xi)\, \Psi(\b q)\, \Theta(\xi-1)\, \frac{d^3q}{q}\, .
\ee
In the derivation of this equation we took into account that
\be
U(\xi,0)=u(\xi)\,\Theta(\xi-1)\,.
\ee
Here $u(\xi)$ is defined by equation~(\ref{jtt23a}), and $\Theta(x)$ is the
Heaviside step function (i.e., $\Theta(x)=1$ if $x>0$ and $\Theta(x)=0$ if
$x<0$).

Obviously, in the case of isotropic turbulence, the general
equation~(\ref{iso11}) should coincide with equation~(\ref{jtt23}).
Moreover, equation~(\ref{jtt23}) can describe even the case of
anisotropic turbulence, if one substitute the function $\Psi(q)$ by
the spectrum of turbulence averaged over directions of the vector $\b
q$, $\Psi(q)\equiv\langle\Psi(\b q)\rangle$. This implies that the
averaged radiation power does not depend on the structure of the
correlation tensor. In particular, the monotonic decrease of the
intensity given by equation~\eqref{jtt23aa}, also is observed in the
case of anisotropic turbulence.  

Although this conclusion is derived under the assumption of isotropic
distribution of emitting particles, in fact the obtained result is
valid also for the case of anisotropic particle distribution when the
change of particle density is small for typical ``angular shifts'' of
the value of $1/\gamma$. Thus, as long as the emission is considered
in the ultrarelativistic regime, this assumption can be broken only in
the case of highly collimated particle beams (which is likely
inconsistent with the underlining assumption of the turbulent magnetic
field). Also we note that if we deal with strongly anisotropic
distribution of particles, the radiation does depend on the structure
of the correlation tensor, therefore it is important to define it
correctly (see Section \ref{sec:prev}).

\section{Comparison with previous results}\label{compare}
\label{sec:prev}
In recent years, a large number of studies have been devoted to
calculations of radiation (the magnetic bremsstrahlung) generated by charged
particles in small-scale turbulent
magnetic fields. However, to our knowledge, the general expression for the
radiation spectrum described by
equation \eqref{jtt23}, is derived for the first time in this paper.
Also, in the previous studies a few additional conditions have been assumed,
which however appear redundant, and
actually not needed at all in the framework of our approach. This redundancy not
only
superficially constraints the applicability of the obtained results, but also
introduces some confusion in the analysis and comparison of different radiation
regimes.
Finally, some solutions and related conclusions derived in this paper do not
coincide with the results of previous studies.
Therefore, we present below a short overview of a few important papers on the
topic, compare their main results
with our study, and outline the key differences between the approaches which
might cause, in
our view, these discrepancies.

There are two basic theoretical approaches to study radiation in random magnetic
fields. The first one is based on the seminal
paper by \citetalias{Toptygin_Fleish}, where a kinetic equation has
been derived for the probability of different
particle trajectories in a chaotic magnetic field (see Eq.(12)$_{\rm TF87}$),
and an
approximate solution has been found to this equation
\citep[see also][for a simplified description of the approach of
\citetalias{Toptygin_Fleish}]{Fleish06a}.
However, the introduced simplifications
significantly limit the applicability of this approach and do not allow a
self-consistent testament of the problem. More
specifically, we discuss these issues bellow.

The second approach is based on the perturbation theory
(\citetalias{Medved00};\citealt{Medved06,Fleish06}).
In all these papers, the authors start from an expression for the emission
produced by
a particle deflected by {\it a small angle} in a magnetic field localized in a
compact
region of space \citep[see][\S~77]{Landau2}. This expression can be written as
\be\label{FM1}
\frac{d\mathcal{E}_{\omega}}{d\omega}=\frac{e^2\omega}{2\pi c^3}\!\!
\int\limits_{\omega/(2\gamma^2)}^\infty\!\!\!\frac{|{}\,\b
a_{\omega'}|^2}{\omega'^2} \left(1-\frac{\omega}{\omega'\gamma^2}
+\frac{\omega^2}{2\omega'^2\gamma^4} \right) d\omega',
\ee
where $\b a_{\omega'}=\int_{-\infty}^\infty\! \b a(t)\,e^{i\omega't}\,dt$ is the
Fourier component of acceleration.
However, if the magnetic field occupies a large volume, then even in the case of
chaotic magnetic field,
the particle deflection will be (unavoidably) large (because of multiple
scatterings over emission correlation length). Therefore, the solution based on
this expression has a rather limited applicability compared to the practical
realizations in the chaotic magnetic field.

The approached employed in our study also is based on the perturbation theory,
but it is valid
when the particle deflection is small on the typical magnetic field correlation
length,
 or, equivalently, if $R_{\rm L}\gg\lambda$.

Note that the later approximation was also implicitly used in
\citetalias{Toptygin_Fleish} (see
Eq.(11)$_{\rm TF87}$) when deriving the kinetic equation. So even in the case of
precise solution of this equation, the
results cannot be expanded beyond the parameter region described by the
perturbation theory approach
presented in our paper.
Moreover, since the derived kinetic equation appeared to be too complex to be
treated analytically,
a few further simplifications have been introduced to obtain an analytical
solution. In particular, the original Eq.(12)$_{\rm TF87}$ was replaced by
Eq.(17)$_{\rm TF87}$, which indeed could be equivalent to the original one if in
the rhs
of this equation they would use the
function $q(\omega,\theta)$ determined by Eq.(15)$_{\rm TF87}$. In Eq.(17)$_{\rm
TF87}$
 $\omega$ enters as a parameter, thus for solution of this equation it can be
taken as a constant, and the function $q$ treated as a function of one variable
$q(\theta)$. However, since in this case the equation does not have a
solution, the authors replace the function $q(\omega,\theta)$ by an empirical
function
$q(\omega)$. This simplification allows an analytical solution, but since it
concerns the
term with the highest derivative in
the equation, the uncertainties imposed by this substitution cannot be evaluated
and correspondingly, the
limits of applicability remain highly unknown. Note that the empirical function
$q(\omega)$ itself determines the radiation spectrum in the case of absence of
the regular component of the magnetic field.
However, within the framework of theory of \citetalias{Toptygin_Fleish} this
function, strictly
speaking, is not derived.
Instead, based on arguments of the asymptotic behavior, they have proposed
the following form $q(\omega)=q(\omega,\theta=\theta_*)$, where
$\theta_*^2=(a-1)/\gamma^2$ and the value of parameter $a$ was determined
``from the requirement
that at high frequencies, where the perturbation expansion (the method of
equivalent photons, see Appendix) is valid, the present method yields the same
result as the perturbation expansion''.

Let's consider now the results of \citet{Fleish_Biet}, where the approach of
\citetalias{Toptygin_Fleish}
has been applied to the case of the random
magnetic field without a regular component. To make the comparison transparent
and less bulky,
discuss the results for the fix value of the index of the turbulence spectrum
$\alpha=2$, and
ignore the impact of the surrounding medium (i.e., assume $\omega_p=0$).
For this specific case, the spectrum obtained in \citet{Fleish_Biet}, can be
expressed as:
\be\label{FB1}
P_\omega=\frac{8e^2\gamma^2}{3\pi c}\,q(\omega)\,\Phi(s)\,.
\ee
Here
\be\label{FB2}
q(\omega)=\frac{\omega_{st}^2\omega_0^{}\gamma^2}{
(a\omega/2)^2+(\omega_0\gamma^2)^2}\,,
\ee
\be
\Phi(s)=24s^2\int\limits_0^\infty\! dt\,e^{-2st}\sin(2st)
\left(\coth t-\frac1t \right),
\ee
where
\be
s=\frac1{8\gamma^2}\left(\frac{\omega}{q(\omega)}\right)^{\!1/2}\,,
\quad \omega_{st}=\frac{c}{R_L}\,,\quad \omega_0^{}=\frac{c}{\lambda} \,.
\ee
$\Phi(s)$ has the following asymptotic limits \citep{Fleish_Biet}:
\be\label{FB2a}
\Phi(s)\approx 1\,,\;\; {\rm if}\; s\gg 1\,;\quad
\Phi(s)\approx 6s\,,\;\; {\rm if}\; s\ll 1\,.
\ee

At $\omega\sim \omega_j=\omega_0^{}\gamma^2$ the $s$ parameter is large,
$s\sim
\omega_0^{}/\omega_{sl}=R_L/\lambda\gg1$. Thus one can use the asymptotic limit
for $\Phi(s)=1$. Then
equation~(\ref{FB1}) can be expressed in a simple form
\be\label{FB3}
P_\omega=\frac{8e^2\gamma^2}{3\pi c}\,q(\omega)\,,
\ee
Apparently, the function $q(\omega)$ determines the shape of the radiation
spectrum.
However, this function has not be derived either by \citet{Fleish_Biet} or by
\citet{Fleish06a}.
We can only guess that the authors have used the simplified form of
Eq.(39)$_{\rm TF87}$ (after removal of the
bulky complex term from that equation).

\citet{Fleish_Biet} performed numerical calculation of the radiation spectra
also for the case
of strong random magnetic field, i.e. in the regime of $\lambda \gtrsim
R_L$. In the asymptotic limit of $\lambda \gg R_L$, the spectrum can be
obtained analytically; in this regime we deal with
 the standard synchrotron spectrum described by equation~(\ref{sy_ra22a}).
 On the other hand, equation~(\ref{FB2a}) with the asymptotic limit of
$\Psi(s)$ for $s\ll1$ from equation~(\ref{FB2a}),
 differs significantly from equation (\ref{sy_ra22a}). In our view, the reason
for this discrepancy is that the basic kinetic equation in the theory of
\citetalias{Toptygin_Fleish} is derived under assumption of $\lambda\ll R_L$
(see
Eq.(9)$_{\rm TF87}$). Thus, this approach cannot be applied for the regime
$\lambda
\gtrsim R_{\rm L}$.

The fact that equation (\ref{FB1}) is not applicable for the case of $\lambda\gg
R_L$ can be also illustrated by computation of the total power emitted by a
particle.
Let's consider the ratio of the radiated and lost energies by the relativistic
charged particle:
\be\label{FB4}
\rho=\int\limits_0^\infty \! P_(\omega)\,d\omega\Big/ I\,.
\ee
Here $I$ and $P_\omega$ are determined by equations (\ref{jtt27}) and
(\ref{FB1}), respectively. For a particle
emitting in vacuum, the condition $\rho=1$ should be satisfied. In the
asymptotic case of $\lambda\ll R_L$, one can use equation (\ref{FB3}) and
demonstrate that for $a=2$ we indeed have $\rho=1$. However,
in the limit of $\lambda\gg R_L$,
\be
\rho=5.6\times p^{2/3}\,,
\ee
where
\be
p= \frac14 \,\sqrt{\frac{3}{2a}}\, \frac{\omega_0^{}}{\omega_{st}}=
\frac14 \,\sqrt{\frac{3}{2a}}\, \frac{R_L}{\lambda}\ll 1\, ,
\ee
i.e. the condition $\rho=1$ is violated.

Thus, we can conclude that in the case when the non-chaotic magnetic field is
nil, the approached developed by \citetalias{Toptygin_Fleish} has a very limited
applicability. Namely, one can derive the spectrum in the form of equation
(\ref{FB3}) with function $q(\omega)$ constrained by asymptotic behavior only.

Now let's compare our results with the studies based, like our paper, on the
perturbation theory.

In \citetalias{Medved00}, a specific geometry of interaction has been
postulated.
Namely, it was assumed that the particle moves
along axis $x$, and that magnetic field has only $y$ component, $B_y$.
Therefore, acceleration is parallel to $z$-axis:
$a(t)=\frac{e}{m\gamma} \,B_y(vt,0,0)\equiv \frac{e}{m\gamma}\,B(t)$. Although
the radiation power obtained in
Sect.~\ref{sec:chaotic} was derived under the assumption of homogeneity of
turbulence, and thus is not
applicable to the case considered by \citetalias{Medved00}, it is
straightforward
to apply our approach to this case also. Namely, accepting the definition of the
spectral power given by equation \eqref{jitt2},
one can average over the magnetic field configurations in
equation~(\ref{FM1}). This gives
\[
P_{\omega}(t)= \frac{e^4\omega}{2\pi m^2\gamma^2 c^3}
\]
\be\label{FM2}
\times\!\!\!\! \int\limits_{\omega/(2\gamma^2)}^\infty\!\!
\frac{(\t B^2)_{\omega'}}{\omega'^2} \left(1-\frac{\omega}{\omega'\gamma^2}
+\frac{\omega^2}{2\omega'^2\gamma^4} \right) d\omega',
\ee
where
\be\label{FM3}
(\t B^2)_{\omega'}\equiv \int\limits_{-\infty}^\infty \!
\langle B(t+\tau/2)B(t-\tau/2)\rangle \,e^{i\omega'\tau}\,d\tau\,.
\ee
is the Fourier component of the magnetic field correlation function; generally
it may depend not only on
$\omega'$, but also on $t$.
Obviously,
\be\label{FM4}
 \langle B^2(t)\rangle = \int\limits_0^\infty \!
(\t B^2)_{\omega}\,\frac{d\omega}{\pi}\,.
\ee

Equation~\eqref{FM2} describes the radiation power for the geometry postulated
by for an arbitrary
spectrum of turbulence. Then, a specific spectrum of turbulence has been
considered in \citetalias{Medved00}
for derivation of Eq.(17)$_{\rm M00}$. The interesting feature of the latter is
that in
the limit of $\omega\rightarrow0$
the spectrum $P_\omega\sim \omega$, and contains an abrupt cutoff, $P_\omega
=0$ at $\omega>2\omega_j$.
However, we should note that these spectral features do not characterize the
jitter radiation in general
(as it can be seen from equation~\eqref{FM2}), but simply are the consequence of
the
choice of a specific turbulence
spectrum and/or interaction geometry \citep[see also][]{Fleish06}. Indeed, if
one adopts a different
turbulence spectrum,
e.g. $(\t B^2)_{\omega'} \sim(\omega'^2+\omega_*^2)^{-\alpha}$,
then for any positive value of the index $\alpha$, the spectrum is a
monotonically decreasing function of $\omega$.
Moreover, if the stochastic field has both $y$- and $z$- components, and the
correlation
function is azimuthally symmetric in respect to $x$-axis, then even for the
power-law spectrum of turbulence, adopted by
\citetalias{Medved00}, the spectrum is not expected to be linear in the limit of
small $\omega$.

The treatment of radiation in a chaotic magnetic field always leads to the
appearance of the
correlation tensor, $\t K_{\rho\sigma}$ (see equation~(\ref{jtt12}) of this
paper and Eq.(12) of
\citealt{Fleish06}). However, often the structure of this tensor, $\t
K_{\rho\sigma}(\b q,\k)$,
is wrongly postulated. If we consider a homogeneous environment
without preferred directions, then we deal with only two second-order tensors:
$\d_{\rho\sigma}$ and $q_\rho q_\sigma$. Therefore, the correlation tensor
should have the following structure: $\t
K_{\rho\sigma}=c_1\,\d_{\rho\sigma}+c_2\,q_\rho q_\sigma$, where $c_{1,2}$ are
two scalar functions. The transversality
condition implies $c_1+c_2 q^2=0$, thus $\t K_{\rho\sigma}$ has to be
proportional to
$(\d_{\rho\sigma}-q_\rho q_\sigma/q^2)$, as used in equation~(\ref{jtt14}).
However,
in the case of existence of a distinct direction, $\b s$, e.g. normal to the
shock front,
the tensor structure becomes more complex:
\be
\t K_{\rho\sigma}=c_1\,\d_{\rho\sigma}+c_2\,q_\rho q_\sigma+c_3\,s_\rho
s_\sigma+ c_4\,q_\rho s_\sigma+c_5\,s_\rho q_\sigma\,.
\ee
The transversality condition imposes three constraints on the
scalar functions $c_i$, thus the correlation tensor $\t K_{\rho\sigma}$ must
have the following structure:
\[
\t K_{\rho\sigma}=\Psi_1\left(\d_{\rho\sigma}-
\frac{q_\rho q_\sigma}{q^2} \right)
\]
\be\label{jtt13b}
+\Psi_2 \left( s_\rho-q_\rho\,\frac{(\b s\b
q)}{q^2}\right)\left( s_\sigma-q_\sigma\,\frac{(\b s\b q)}{q^2}\right),
\ee
where functions $\Psi_{1,2}$ depend on three arguments: $|\b q|$, $(\b s\b q)$
and $\k$.
In a gyrotropic medium,
the correlation function may contain an additional term:
$\e_{\rho\sigma\tau}q_\tau\,\Psi_3$, where $\e_{\rho\sigma\tau}$ is the
Levi-Civita-tensor and $\Psi_3$ is a complex function as it follows from the
general theory of
fluctuations \citep[see e.g.][\S~122]{Landau5}. Note, however,
that this term does not contribute to the emission power, since in
equation~(\ref{jtt16}) the tensor $\t K_{\rho\sigma}$ is convolved with a
symmetric tensor.

For the additional assumption that the magnetic field is perpendicular to the
direction $\b s$,
the correlation function should satisfy the equations $\t
K_{\rho\sigma}s_\rho=0$, $\t K_{\rho\sigma}s_\sigma=0$. In this case the
functions $\Psi_1$ are linked $\Psi_2$ via the relation
\be
\Psi_1+(1-(\b s\b q)^2/q^2)\,\Psi_2=0\, ,
\ee
and the correlation function $\t K_{\rho\sigma}$ is determined just by one
scalar function.

However, in some studies dealing with the anisotropic turbulence, different
tensor structures have been proposed
for the correlation function: $\t K_{\rho\sigma}\propto (\d_{\rho\sigma}-s_\rho
s_\sigma)$ - we can refer, for example to
Eq.(8) in \citealt{Medved06}, Eq.(18) in \citealt{Fleish06}, Eq.(10) in
\citealt{Medved11}, Eq.(11) in \citealt{Medved12}). This correlation function
does not satisfy the transversality condition, i.e., the considered magnetic
field is not divergence free, $\nabla\b B\neq0 $. Apparently, this is a wrong
result,
therefore the results obtained under the assumption of $\t
K_{\rho\sigma}\propto (\d_{\rho\sigma}-s_\rho s_\sigma)$ should be revised.

It is important to note that certain mathematical operation
often used for computation of emission in chaotic magnetic field
lack mathematical strictness \citep[also see discussions
in][]{Fleish06,medeved05}. In particular, this concerns the involvement of
the field harmonics, $B_k$, which implies that the Fourier
transformation can be applied to the stochastic magnetic field. This
assumption hardly can be justified or disproved from the first
principles, however this assumption may lead to a rather controversial
expression for the Fourier image of the correlation function.
For example, the following structure has been obtained for the correlation
tensor $\t K_{\rho\sigma}\propto
\t B_\rho\t B_\sigma$ \citep[see footnote 2 and equation 5
in][respectively]{Fleish06,Medved06}, which, however, contradicts the general
tensor structure given by equation~\eqref{jtt13b}\citep[see
also][\S~122]{Landau5}.

Finally, we note that in the framework of our
approach, no assumptions regarding the properties of the stochastic
magnetic field are required. Instead, we assumed that the Fourier
transformation can be applied to the magnetic field {\it correlation
function}, which is a significantly less demanding assumption.

\section{Discussion and Summary}
\label{sec:conc}

The so-called jitter radiation mechanism represents a version of
the magnetic bremsstrahlung of relativistic charged particles
in a turbulent magnetic field. This regime of radiation can be realized with
an efficiency as high as the ``standard'' synchrotron radiation but with quite
distinct energy spectrum strongly shifted towards higher energies. This
makes the jitter radiation an attractive gamma-ray production channel in
highly turbulent astrophysical environments.

 In this paper we present a novel study on spectral properties of
the jitter radiation performed within the framework of perturbation theory
in the regime of the so-called small-scale turbulence, when the coherent
length of the field is significantly
smaller than the
nonrelativistic Larmor radius, $\lambda \ll R_L=mc^2/eB$, or

\be\label{scale}
\lambda \ll 1.7 \times 10^3 (m/m_e) (B/1 \, \rm G)^{-1} \, \rm cm \, .
\ee
Here $B$ is the average magnetic field, and $m$ is the mass of radiating
charged particle. It is remarkable that the condition for realization of the
jitter regime does not depend on the particle
energy, but only on its mass. For example, for electrons the condition imposed
by equation (\ref{scale}) implies a turbulence scale as small as 100km in young
supernova remnants, less than 10m in gamma-ray production regions of blazers,
and
1cm in GRBs, assuming typical values of the magnetic field in these objects of
about $100\mu G$, $1 \rm G$ and $1 \rm kG$, respectively. For protons these
conditions are relaxed by three orders of magnitude. However,
the magnetic bremsstrahlung of protons is a much slower process compared to
electrons. It becomes adequately effective only at very high energies of protons
and at the presence of large magnetic field which in its turn implies tighter
conditions on the turbulence scale. Whether turbulent fields can be generated at
scales imposed by equation (\ref{scale}) is a non-trivial issue the discussion
of which is beyond the scope of this paper. Here we focused merely on the study
of radiation properties and perform calculations under the assumption that
equation~(\ref{scale}) is (by definition) fulfilled. We derived
an expression for the spectral power of radiation presented in
a general but rather compact form convenient for numerical calculations.

The jitter radiation has a simple spectral form. Its SED for a monoenergetic
particle distribution is shown
in Fig.\ref{fig:jitt} together with the SED of synchrotron radiation. Both SEDs
have pronounced maximums separated from each other by a factor of $R_L/\lambda$.
Obviously when the jitter regime is realized, the maximum of its SED is shifted
towards higher energies (the position of the peak in the synchrotron SED is at
the energy $\approx 1.115 \omega_c$). Unless one introduces strong assumptions
regarding the turbulence spectrum and/or its geometry, the low energy part of
the spectrum has standard photon index $\Gamma=1$. It is hard
but still softer than the spectrum of synchrotron radiation with $\Gamma=4/3$.
The jitter and synchrotron spectra are very different beyond their respective
maximums. While the standard synchrotron spectrum cuts off quite sharply (exponentially)
just after the maximum, the spectrum of the jitter radiation continues as a
power-law until the energy $\sim (R_L/\lambda)^3 \omega_c$ with a photon index
$\Gamma=\alpha+1$, where $\alpha$ is the power-law index of the
turbulence spectrum (in the framework of the perturbation theory,
the spectral shape of radiation above this limit cannot be calculated).
Remarkably, this part of the spectrum is not sensitive
to the details of the energy distribution of particles, but depends only on
the position of the cutoff in the particles distribution. The latter determines
 the start of the power-law tail which
should be (by definition) quite long since $R _L \gg \lambda$.
For example, if the ratio $R_L/\lambda$ exceeds 10,
the power-law tail of the jitter radiation, which mimic the turbulence spectrum,
would span over more than two energy decades after the maximum. Bellow the
maximum, the jitter radiation does depend on the particle distribution. In
particular, if the relativistic particles have a
power-law distribution with an index $\mu$, the spectrum of the jitter
radiation is also power-law with photon index
 $\Gamma=(\mu + 1)/2$, i.e. exactly the same as in the case of synchrotron
radiation.

In this paper we do not aim to discuss astrophysical implications of jitter
radiation which deserve separate consideration. Instead, we would rather comment
on some interesting features which make this mechanism quite unique amongst
other high energy radiation processes.

First of all, the slight dependence (or rather independence) of the high
energy spectral tail on the distribution of parent relativistic particles,
is quite unusual and apparently does not have an analog in astrophysics.
Moreover, the fact that the spectral shape of
this tail simply mimics the spectrum of turbulence, opens a unique opportunity
for the straightforward probe of the spectrum of small-scale turbulence by
measuring the characteristic high energy electromagnetic radiation
and identifying it with the jitter mechanism.

While in the case of injection of relativistic electrons into a highly turbulent
medium, where the condition of equation (\ref{scale}) is satisfied,
guaranties production of radiation in the jitter regime, the questions of its
detection depends on the total energetics in relativistic particles. Given the
usually (very) high efficiency of jitter radiation,
and for typical parameters characterizing the
nonthermal astrophysical sources of both galactic and extragalactic origin, the
 production of detectable fluxes of jitter radiation in the X-ray and/or
gamma-ray bands could be readily realized in standard acceleration and radiation
scenarios.

The identification of the origin or radiation is the second critical issue.
Fortunately, the jitter radiation does provide us with distinct features which
can be used to identify its nature. In particular, for a standard power-law
injection distribution of electrons with a high energy cutoff given, for
example, in the form of equation~(\ref{pl_cutoff}) with $\mu=2$, and assuming
a Kolmogorov-type spectrum of turbulence, $\alpha=5/3$, we expect a gamma-ray
spectrum which can be described as broken power-law. The high energy part of the
spectrum above the break is expected to have a photon index of
$\Gamma_2=\alpha+1\simeq 2.7$, while the low energy part (below the break)
$\Gamma_1=(\mu+1)/2=1.5$ or $\Gamma_1=(\mu+2)/2=2$ for the slow and fast
cooling regimes, respectively. This corresponds to the change of the spectral
index by $\Delta \Gamma=1.2$ or $0.7$ depending on the cooling regime.
Such a behavior differs significantly from the standard synchrotron cooling
break with $\Delta \Gamma=0.5$. In the case of a low energy cutoff in the
electron spectrum, which is often required to fit the data, e.g.
the spectra of gamma-ray blazars, we should expect another break below which the
photon index would be $\Gamma=1$ Therefore, in the case of detection of a
non-standard broken power-law spectra,
especially when the high energy power-law tail has a photon index close to
2.5 and extends well beyond the break, the jitter mechanism can be be treated
as a process responsible for the observed spectral features \citep[see
also][]{Fleish_Biet}.

Despite all attractive properties of synchrotron radiation of ultrarelativistic
electron, its spectrum usually terminates before reaching the gamma-ray
domain. Even in the extreme accelerators it cannot extend beyond the so-called
synchrotron limit $\sim 100$~MeV, unless being additionally Doppler boosted in
sources with relativistic Doppler factors. This can be the case, for example,
of the recently discovered flares of the Crab Nebula \citep[see, e.g.][and
references therein]{Rolf,crab_flare} or the multi-GeV counterparts of gamma-ray
bursts \citep{fermi_grb}.
On the other hand, the jitter mechanism may offer another possibility for
extension of the spectrum well beyond the
energy synchrotron limit. We should note in this regard that in the case of
fulfillment of the condition in
equation~(\ref{scale}), the appearance of
jitter radiation is not only unavoidable, but its spectrum could extend to
high or even very high energies. A more
a principal issue in this regard is the challenge of formation of turbulence
on very small scales, e.g. through the
Weibel type instabilities.

\appendix
\section{High energy asymptotics  of  synchrotron radiation}\label{app_a}

At high frequencies, the intensity of synchrotron radiation decreases exponentially,  $G(x)\propto
\exp\!\left(-\frac{2\omega mc}{3eB\gamma^2}\right)=\exp\!\left(-\frac{\omega}{b\,\omega_0}\right)$ (see
equation~(\ref{un6x})).  Therefore  at  $\omega\gg \omega_0$  the contribution to the  radiation 
spectrum given by  equation~(\ref{diff_b7}) is dominated by   regions characterized by large magnetic field: $b\gg1$. 
Let us assumed that for $b\gg 1$, the magnetic field strength is distributed as  
power-law : $w\,dB\approx A\,b^{-\sigma}db$.

It is convenient   to start  the calculations of the spectrum from equation~(\ref{sy_ra22}), where magnetic field strength is replaced by  $B\to B\sin\chi$, i.e. before integration over directions of the magnetic field.  This gives
\be\label{appj1}
P_\omega=\frac{\sqrt{3}\,e^2}{2\pi R_{L0}}\,x\int\limits_{x/b\sin\chi}^\infty
\! K_{5/3}(u)\,du\,,
\ee
where $x=\omega/\omega_0$, $R_{L0}=mc^2/eB_0$  and  $b=B/B_0$. To average over the strength of the magnetic field strength, one should 
multiply equation~(\ref{appj1})  to  the distribution of magnetic field, $A\,b^{-\sigma}db$, and then integrate 
over $db$.  Let us introduce  a new variable  $b\to \xi=x/b\sin\chi$, and change   the order of integration over $d\xi$ and $du$. Then,  
after a rather simple analytical integration over $d\xi$,  we obtain:
\be\label{appj2}
\int\!P_\omega w(b)\,db=A\,\frac{\sqrt{3}\,e^2}{2\pi R_{L0}}\,
\frac{x^{-\sigma+2}(\sin\chi)^{\sigma-1}}{\sigma-1}
\int\limits_0^\infty\! u^{\sigma-1} \! K_{5/3}(u)\,du\,.
\ee
The remaining  integral over  $du$ can be expressed in terms of Gamma functions:
\be
\int\limits_0^\infty\!u^{\sigma-1}\,K_{5/3}(u)\,du=
2^{\sigma-2}\,\Gamma\!\left(\frac\sigma2+\frac56\right)
\Gamma\!\left(\frac\sigma2-\frac56\right).
\ee
The integration over pitch-angles also leads to  an expression containing Gamma functions:
\be
\langle(\sin\chi)^{\sigma-1}\rangle=
\frac{\sqrt{\pi}\,\Gamma\!\left(\frac\sigma2+\frac12\right)}
{2\,\Gamma\!\left(\frac\sigma2+1\right)}\,.
\ee
Thus, in the limit of large  frequency $\omega\gg\omega_0$, 
the spectrum of synchrotron radiation averaged over the directions  and strength of the magnetic field, 
is described by a power-law function:
\be\label{appj3}
\langle P_\omega\rangle=A\,\frac{e^2}{R_{L0}}\,C_\sigma\,x^{-\sigma+2}\,,
\ee
where
\be\label{appj4}
C_\sigma=\frac{\sqrt{3}\,2^{\sigma-4}}{\sqrt{\pi}\,(\sigma-1)}\,
\frac{\Gamma\!\left(\frac\sigma2+\frac56\right)
\Gamma\!\left(\frac\sigma2-\frac56\right)
\Gamma\!\left(\frac\sigma2+\frac12\right)}{\Gamma\!\left(\frac\sigma2+1\right)}\,.
\ee
Note that  if  the magnetic filed is formally distributed as pure power-law, $B$: $w(B)=\bar A\, B^{-\sigma}$, 
equation~(\ref{appj3}) gives precise solution  for the  radiation  spectrum. In this case $A=\bar A\,B_0^{1-\sigma}$, and 
therefore $\langle P_\omega\rangle$ appears to be  independent of  $B_0$. 

To understand  the condition for  applicability of  equation~(\ref{appj3}), 
let us  estimate the correction terms to this equation  for a specific distributions of the 
magnetic field. Let's assume, for example,  the following distribution:
\be\label{appj5}
w(B)\,dB=\frac{A\,b^2}{(1+b^2)^{1+\sigma/2}}.
\ee
In the limit  $b\gg 1$, the first two terms of  the series are
\be\label{appj5}
\frac{A\,b^2}{(1+b^2)^{1+\sigma/2}}\approx A\left(\frac1{b^\sigma}
-\frac{1+\sigma/2}{b^{\sigma+2}}\right).
\ee
Correspondingly,  
\be\label{appj6}
\langle P_\omega\rangle=A\,\frac{e^2}{R_{L0}}\left(C_\sigma\,x^{-\sigma+2}-
(1+\sigma/2)\,C_{\sigma+2}\,x^{-\sigma}\right).
\ee
The ratio of these two terms can be expressed as
\be\label{appj7}
r=\frac{(\sigma-1)(9\sigma^2-25)}{18\,x^2}\sim \frac{\sigma^3}{2\,x^2}\,,
\ee
which  implies that one can neglect the second term when $x\gtrsim \sigma^{3/2}$. This 
can be treated as the   condition for applicability of equation~(\ref{appj3}).

\section{The case of anisotropic turbulence}
\label{app_b}

Here we present some intermediate calculations required for derivation of
equation~\eqref{iso8}
from equation~\eqref{iso7}. To compute the integrals, it is convenient to
introduce the following new variables:
\be
\zeta=\gamma^2\theta^2\,,\quad x=\cos\vartheta\,,
\ee
where $\vartheta$ is the angle between vectors $\b\nu$ and $\b q$. The
integration over the azimuthal angle is trivial; 
it gives $d\Omega\,d\Omega_\alpha=\frac{2\pi^2}{\gamma^2}\,dx\,d\zeta$. Then,
the integration of equation~\eqref{iso7}
results in
\be\label{app1}
F_1=\frac{\pi^2}{2}\! \int\limits_0^\infty\! d\zeta\int\limits_{-1}^1\!
dx \! \left(1+x^2\right)\frac{1+\zeta^2}{(1+\zeta)^4}
\,\d(\t\omega+cqx-\k)\,.
\ee
For the upper limit of integration over $\zeta$ we set $\infty$, which is
valid only in the ultrarelativistic regime (see also the discussion after
equation~(\ref{jtt22})). The argument of the $\d$-function nulls for
$x=x_0^{}=(\k-\t\omega)/cq$.
The integral becomes zero if $x_0^{}$ lies beyond the integration interval,
$x_0^{}>1$ or $x_0^{}<-1$. For the value of $x_0^{}$ within the integration
range, i.e., $|x_0^{}|<1$, we obtain
 \be\label{app2}
 F_1=\frac{\pi^2}{2cq}\int\limits_{\zeta_1}^{\zeta_2}\! d\zeta
\left(1+x_0^2\right)\frac{1+\zeta^2}{(1+\zeta)^4}\,,
 \ee
where the lower and upper integration limits, $\zeta_{1,2}$, are determined by
the conditions $|x_0^{}|=1$ and $\zeta\ge 0$.

It is convenient to express these limits as $\zeta_1=\max(0,\xi(\kappa-1)-1)$
and $\zeta_2=\xi(\kappa+1)-1$ ($\zeta_2$ should be positive), where
 \be\label{app3}
\kappa=\frac{\k}{qc}\,, \qquad \xi=\frac{2qc\gamma^2}{\omega}\,.
 \ee
This allows derivation of equation~(\ref{iso8}) via the analytical
integration:
 \be\label{app4}
U(\xi,\kappa)=\Theta\big(\xi(\kappa+1)-1\big)\left[U_1\,
\Theta\big(1-\xi(\kappa-1)\big)
+U_2\,\Theta\big(\xi(\kappa-1)-1\big) \right],
 \ee
where
 \be\label{app5}
 U_1=\frac1{\xi^3}\,\big(\xi(\kappa+1)-1\big) \left(\xi^2(\kappa-1)+4\xi+
 \frac{2\xi^2-2\xi+1}{\kappa+1}-\frac{\xi-1}{(\kappa+1)^2}
 +\frac{2}{(\kappa+1)^3} \right)
 -\frac{3}{\xi^2}\,(\xi\kappa+1)\ln\big(\xi(\kappa+1)\big)\,,
 \ee
and
 \be\label{app6}
 U_2=\frac{2}{\xi^3(\kappa^2-1)^3}\left(
 4+3\kappa^6\xi^2+3\xi\kappa^5-6\xi^2\kappa^4-12\xi\kappa^3+\left(3\xi^2+4\right
) \kappa^2+9\xi\kappa\right)-\frac{3}{\xi^2}\,(\xi\kappa+1)
 \ln\left(\frac{\kappa+1}{\kappa-1}\right).
 \ee
Equation~(\ref{app4}) implies that function $U(\xi,\kappa)$ has non-zero values
only if
 $\kappa>\frac1\xi-1$. The two terms in equation~(\ref{app4}), $U_{1,2}$,
 give non-zero contribution for $\frac1\xi-1<\kappa<\frac1\xi+1$ and
$\kappa>\frac1\xi+1$, respectively. The continues
 function $U$ has a break at $\kappa=\frac1\xi+1$ $U(\xi,\kappa)$.

\section*{Acknowledgments}
We would like to thank A.~Bykov, G.~Fleishman, B.~Reville and
Y.~Teraki for helpful discussions.  We should also mention that during
our communications G.~Fleishman informed us that some of our critical
comments from Sect.~\ref{compare} of this paper concerning the
previous works he had discussed in his book ``Stohasticheskaja teorija
izluchenija'' (in Russian, 2008, ISBN 978-5-93972-624-5).


\end{document}